\documentclass[11pt,a4paper]{article}
\pdfoutput=1

\usepackage{jcappub}
\usepackage{mathtools,bm,booktabs,array,microtype}
\usepackage[T1]{fontenc}
\usepackage{lmodern}
\DeclareMathSizes{8.75}{8.75}{6.125}{5}
\usepackage{adjustbox}
\usepackage{tabularx}
\usepackage{tikz}
\usepackage[nosfdefault]{sourcesanspro}
\usetikzlibrary{arrows.meta,positioning,fit,backgrounds,calc,matrix}

\allowdisplaybreaks[2]
\numberwithin{equation}{section}
\newcommand{\Mpl}{M_{\rm Pl}}
\newcommand{\HI}{H_I}
\newcommand{\fa}{f_a}
\newcommand{\fI}{f_I}
\newcommand{\thetai}{\theta_i}
\newcommand{\OmegaDM}{\Omega_{\rm cdm}}
\newcommand{\Omegaa}{\Omega_a}
\newcommand{\PRR}{\mathcal P_{\mathcal R}}
\newcommand{\PII}{\mathcal P_{II}}
\newcommand{\As}{A_s}
\newcommand{\betaiso}{\beta_{\rm iso}}
\newcommand{\GeV}{\,\mathrm{GeV}}
\newcommand{\dd}{\mathrm d}

\newcommand{\Gresp}{\mathcal G}
\newcommand{\Ttheta}{\mathcal T_{\theta}}
\newcommand{\Ns}{N_*}

\newcolumntype{L}[1]{>{\raggedright\arraybackslash}p{#1}}

\newcommand{\papertitle}{Axion isocurvature and the model building problem
of low-scale inflation}
\newcommand{\paperauthor}{Sarunas Verner}
\newcommand{\paperaffiliation}{Kavli Institute for Cosmological Physics, University of Chicago, 5640 South Ellis Ave.,
Chicago, IL 60637, USA}
\newcommand{\paperemail}{verner@uchicago.edu}
\newcommand{\paperkeywords}{axions, inflation, isocurvature perturbations,
CMB}
\hypersetup{
  pdftitle={Axion isocurvature and the model-building problem of low-scale inflation},
  pdfauthor={Sarunas Verner},
  pdfsubject={Cosmology and astroparticle physics},
  pdfkeywords={axions, inflation, isocurvature perturbations, CMB, reheating}
}
\newcommand{\paperabstract}
{
The pre-inflationary QCD axion is often said to require low-scale inflation.
We derive a general isocurvature bound for a light axion with a
scale-invariant spectrum, treating its inflationary normalization $f_I$
independently of the late-time decay constant $f_a$ and allowing for an
arbitrary axion dark matter fraction, nontrivial perturbation transfer, and
the full anharmonic abundance response. In the minimal benchmark, where
axions constitute all of the dark matter, $f_I=f_a\leq M_{\rm Pl}$, and the
angular perturbation is conserved, the least restrictive CMB limit considered
gives the 95\% C.L. upper bound
$H_I<1.25\times10^{10}\,\mathrm{GeV}$ on the inflationary Hubble scale. For an
initial misalignment angle $\theta_i=1$, the bound strengthens to
$H_I<2.2\times10^7\,\mathrm{GeV}$, with still stronger constraints near the
hilltop. Combining the isocurvature and tensor spectra yields axion--tensor
and axion-conditioned Lyth bounds. If the observed curvature perturbation is
generated by a canonical cold single-field slow-roll inflaton and the tensor
amplitude varies slowly across the observable CMB window, these relations
limit the inflaton excursion to
$\Delta\phi_{\rm CMB}/M_{\rm Pl}\lesssim10^{-7}\text{--}10^{-4}$ and require
$|M_{\rm Pl}V'/V|\sim10^{-8}\text{--}10^{-5}$ while
$M_{\rm Pl}^2V''/V\sim-10^{-2}$. This hierarchy favors models with independent
control of the inflationary scale, potential derivatives, exit, and
reheating, including hybrid, running-mass, and inflection-point models.
Reheating and the requirement of Peccei--Quinn non-restoration sharpen these
conditions, while nonminimal scenarios can relax them by modifying the
primordial fluctuation, its transfer, or the relic abundance.
}

\title{\papertitle}
\author[a]{\paperauthor}
\affiliation[a]{\paperaffiliation}
\emailAdd{\paperemail}
\abstract{\paperabstract}

\begin{document}
\maketitle

\section{Introduction}
\label{sec:introduction}
If the Peccei--Quinn (PQ) symmetry~\cite{Peccei:1977hh,Peccei:1977ur} is
broken before the observable modes exit the horizon and is not restored
thereafter, inflation homogenizes the misalignment angle of the associated
axion~\cite{Weinberg:1977ma,Wilczek:1977pj} across our observable patch and
dilutes pre-existing strings and domain walls. If the axion is light compared
with the inflationary Hubble scale, however, its angular field acquires
nearly scale-invariant vacuum fluctuations. When the axion later contributes
to dark matter, these fluctuations source a cold dark matter (CDM)
isocurvature mode~\cite{Axenides:1983hj,Seckel:1985tj,Linde:1985yf,
Lyth:1989pb,Turner:1990uz,Lyth:1991ub,Lyth:1992tw}. CMB limits on such a mode
therefore provide one of the strongest connections between QCD axion dark
matter and inflation~\cite{Planck:2018vyg,Marsh:2015xka,    
DiLuzio:2020wdo}.

This connection is commonly summarized by saying that pre-inflationary axion
dark matter requires low-scale inflation, with the benchmark
$\HI\lesssim10^7\GeV$ often quoted for an order-one initial misalignment
angle~\cite{Lyth:1991ub,Marsh:2015xka}. Such a statement is meaningful only
after specifying the axion abundance, its canonical normalization during
inflation, the primordial isocurvature spectrum and its correlation with the
adiabatic mode, and the subsequent cosmological evolution. The familiar
estimate is therefore not a universal upper bound. In particular, the
inflationary normalization need not coincide with the late-time decay
constant, the axion need not constitute all of the dark matter, and the
anharmonic response of the relic abundance can substantially strengthen the
constraint near the hilltop.

The implications extend beyond the numerical value of $\HI$. If the observed
curvature perturbation is generated by a canonical cold single-field
slow-roll inflaton, lowering $\HI$ while preserving the measured scalar
amplitude forces the first potential slow-roll parameter $\epsilon_V$ to
become extraordinarily small. Because $\epsilon_V$ is then negligible, the
observed red scalar tilt requires
$\eta_V\simeq(n_s-1)/2\sim-10^{-2}$. The inflationary potential must
therefore have an extremely small local slope while retaining appreciable
negative curvature. We refer to this separation as the
\emph{axion-conditioned inflationary hierarchy}. It converts the axion
isocurvature limit into a requirement on the shape, radiative stability, and
ultraviolet completion of the inflationary potential, rather than merely on
its overall energy scale.

A related consequence follows from the primordial tensor spectrum, whose
connection to axion isocurvature provides a direct test of axion dark
matter~\cite{Marsh:2014qoa}. The ordinary Lyth argument~\cite{Lyth:1996im,Efstathiou:2005tq} converts a tensor signal into a lower
bound on the inflaton displacement. In the present setting, the axion
isocurvature ceiling instead bounds the inflationary Hubble scale and,
assuming the standard vacuum tensor spectrum in Einstein gravity, the tensor
amplitude. For a canonical inflaton and a tensor amplitude that does not vary rapidly across the observable CMB window, this gives an axion-conditioned upper bound on the corresponding field excursion. The pre-inflationary axion
therefore constrains both the inflationary energy scale and the local
field-space distance traversed while observable modes leave the horizon.

Current CMB measurements sharpen this model-building problem in two logically distinct ways. A dedicated analysis of an uncorrelated, scale-invariant CDM isocurvature mode finds that adding ACT or SPT data does not tighten the amplitude bound relative to Planck alone. For the scale-invariant case, the Planck+ACT bound is mildly weaker because of parameter correlations~\cite{Petretti:2026ayw}. ACT nevertheless affects the inflationary interpretation because commonly used ACT DR6 combinations prefer a somewhat larger scalar spectral index than Planck alone~\cite{AtacamaCosmologyTelescope:2025blo}, while SPT-3G provides independent small-scale measurements consistent with Planck and ACT DR6~\cite{SPT-3G:2025bzu}. Thus, the small-scale data do not presently lower the direct scale-invariant isocurvature ceiling, but the larger ACT-preferred value of $n_s$ makes many low-scale plateau and hilltop realizations more difficult.

Reheating can sharpen this tension. For the post-inflationary histories
considered here, lowering the inflationary scale typically reduces the number of $e$-folds $N_*$ between CMB horizon exit and the end of inflation~\cite{Liddle:2003as,Planck:2018jri}. Requiring the PQ symmetry to remain
broken can also bound the reheating and maximum plasma temperatures and
constrain couplings between the inflaton and PQ sectors~\cite{Kofman:1995fi,Harigaya:2015hha}. When these conditions necessitate a
delayed, matter-like reheating phase, the lower reheating temperature further reduces $N_*$ and shifts standard plateau and hilltop predictions toward a redder scalar spectrum. The minimal scenario therefore presents a three-way model-building tension: isocurvature favors a very low inflationary scale, the observed tilt requires non-negligible negative curvature, and PQ non-restoration can restrict the thermal history in a direction that makes common low-scale predictions still redder.

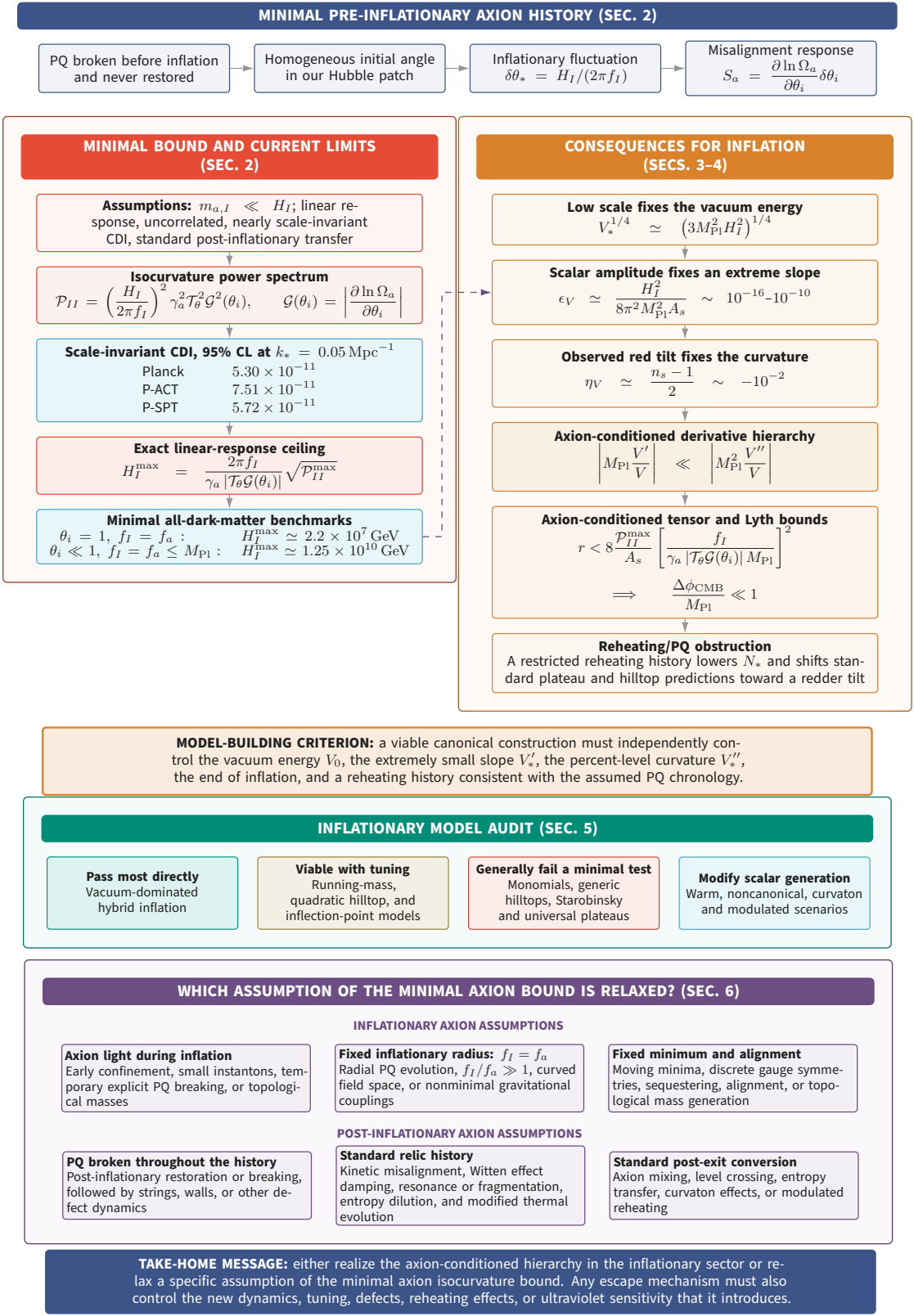
\begin{figure}[p]
\centering
\definecolor{roadmapblue}{HTML}{3C5488}
\definecolor{roadmapred}{HTML}{E64B35}
\definecolor{roadmaporange}{HTML}{D9822B}
\definecolor{roadmapcyan}{HTML}{4DBBD5}
\definecolor{roadmapgreen}{HTML}{00A087}
\definecolor{roadmapviolet}{HTML}{76558F}
\definecolor{roadmapgold}{HTML}{A58B3D}
\begin{adjustbox}{max totalsize={0.98\textwidth}{0.94\textheight},center}
\begin{tikzpicture}[
  scale=0.80,
  transform shape,
  x=1mm,
  y=1mm,
  font=\fontfamily{SourceSansPro-TLF}\selectfont\fontsize{8.75}{10.0}\selectfont,
  text=black!88,
  >=Latex,
  line width=0.45pt,
  soft/.style={
    -{Latex[length=1.8mm,width=1.1mm]},
    draw=black!42,
    line width=0.45pt
  },
  link/.style={
    -{Latex[length=1.8mm,width=1.1mm]},
    draw=roadmapviolet!88!black,
    dashed,
    line width=0.5pt
  },
  box/.style={
    rounded corners=1.6pt,
    draw,
    align=center,
    inner sep=2.8pt,
    line width=0.48pt,
    fill=white
  },
  head/.style={
    rounded corners=1.8pt,
    draw,
    line width=0.78pt,
    align=center,
    inner sep=3.1pt,
    font=\fontfamily{SourceSansPro-TLF}\selectfont\bfseries\fontsize{9.8}{11.1}\selectfont
  },
  rowlabel/.style={
    align=center,
    text=roadmapviolet!88!black,
    font=\fontfamily{SourceSansPro-TLF}\selectfont\bfseries\fontsize{7.7}{8.7}\selectfont
  },
  smallbox/.style={
    rounded corners=1.5pt,
    draw,
    align=center,
    inner sep=2.2pt,
    line width=0.48pt,
    minimum height=10mm,
    fill=white,
    font=\fontfamily{SourceSansPro-TLF}\selectfont\mdseries\fontsize{8.75}{10.0}\selectfont
  },
  escape/.style={
    rounded corners=1.5pt,
    draw=roadmapviolet!88!black,
    fill=roadmapviolet!9,
    align=left,
    inner sep=2.2pt,
    line width=0.48pt,
    text width=50mm,
    minimum height=14mm,
    font=\fontfamily{SourceSansPro-TLF}\selectfont\fontsize{8.0}{9.1}\selectfont
  },
  modelgood/.style={
    rounded corners=1.5pt,
    draw=roadmapgreen!90!black,
    fill=roadmapgreen!13,
    align=center,
    inner sep=2.2pt,
    text width=38mm,
    minimum height=15mm,
    font=\fontfamily{SourceSansPro-TLF}\selectfont\fontsize{8.0}{9.1}\selectfont
  },
  modelcond/.style={
    rounded corners=1.5pt,
    draw=roadmapgold!92!black,
    fill=roadmapgold!14,
    align=center,
    inner sep=2.2pt,
    text width=38mm,
    minimum height=15mm,
    font=\fontfamily{SourceSansPro-TLF}\selectfont\fontsize{8.0}{9.1}\selectfont
  },
  modelbad/.style={
    rounded corners=1.5pt,
    draw=roadmapred!90!black,
    fill=roadmapred!10,
    align=center,
    inner sep=2.2pt,
    text width=38mm,
    minimum height=15mm,
    font=\fontfamily{SourceSansPro-TLF}\selectfont\fontsize{8.0}{9.1}\selectfont
  },
  modelbeyond/.style={
    rounded corners=1.5pt,
    draw=roadmapcyan!88!black,
    fill=roadmapcyan!13,
    align=center,
    inner sep=2.2pt,
    text width=38mm,
    minimum height=15mm,
    font=\fontfamily{SourceSansPro-TLF}\selectfont\fontsize{8.0}{9.1}\selectfont
  }
]


\node[
  head,
  draw=roadmapblue,
  fill=roadmapblue,
  text=white,
  text width=180mm
] (tophead) at (0,0)
{MINIMAL PRE-INFLATIONARY AXION HISTORY (SEC.~2)};

\node[
  smallbox,
  draw=roadmapblue,
  fill=roadmapblue!7,
  text width=38mm,
  below=3mm of tophead,
  xshift=-67mm
] (T1)
{PQ broken before inflation\\and never restored};

\node[
  smallbox,
  draw=roadmapblue,
  fill=roadmapblue!7,
  text width=38mm,
  right=5mm of T1
] (T2)
{Homogeneous initial angle\\in our Hubble patch};

\node[
  smallbox,
  draw=roadmapblue,
  fill=roadmapblue!7,
  text width=38mm,
  right=5mm of T2
] (T3)
{Inflationary fluctuation\\
{\mathversion{normal}$\displaystyle
\delta\theta_*=H_I/(2\pi f_I)$}};

\node[
  smallbox,
  draw=roadmapblue,
  fill=roadmapblue!7,
  text width=38mm,
  right=5mm of T3
] (T4)
{Misalignment response\\
$\displaystyle
S_a=
\frac{\partial\ln\Omega_a}{\partial\theta_i}
\delta\theta_i$};

\foreach \u/\v in {T1/T2,T2/T3,T3/T4}
  \draw[soft] (\u)--(\v);


\coordinate (branchcenter) at (tophead.center |- T4.south);

\node[
  head,
  draw=roadmapred,
  fill=roadmapred,
  text=white,
  text width=84mm,
  anchor=north east
] (LH) at ([xshift=-4mm,yshift=-8mm]branchcenter)
{MINIMAL BOUND AND CURRENT LIMITS\\(SEC.~2)};

\node[
  head,
  draw=roadmaporange,
  fill=roadmaporange,
  text=white,
  text width=84mm,
  anchor=north west
] (RH) at ([xshift=4mm,yshift=-8mm]branchcenter)
{CONSEQUENCES FOR INFLATION\\(SECS.~3--4)};


\node[
  box,
  draw=roadmapred,
  fill=roadmapred!6,
  text width=78mm,
  below=3mm of LH
] (L1)
{\textbf{Assumptions:}
$m_{a,I}\ll H_I$; linear response, uncorrelated, nearly scale-invariant CDI,
standard post-inflationary transfer};

\node[
  box,
  draw=roadmapred,
  fill=roadmapred!8,
  text width=78mm,
  below=3mm of L1
] (L2)
{\textbf{Isocurvature power spectrum}\\[-1pt]
$\displaystyle
\mathcal P_{II}
=
\left(\frac{H_I}{2\pi f_I}\right)^2
\gamma_a^2\mathcal T_\theta^2\mathcal G^2(\theta_i),
\qquad
\mathcal G(\theta_i)
=
\left|
\frac{\partial\ln\Omega_a}{\partial\theta_i}
\right|$};

\node[
  box,
  draw=roadmapcyan!90!black,
  fill=roadmapcyan!13,
  text width=78mm,
  below=3mm of L2
] (L3)
{\textbf{Scale-invariant CDI, 95\% CL at
$k_*=0.05\,\mathrm{Mpc}^{-1}$}\\[1pt]
\begin{tabular}{@{}l@{\hspace{10mm}}r@{}}
Planck     & $5.30\times10^{-11}$\\
P-ACT & $7.51\times10^{-11}$\\
P-SPT & $5.72\times10^{-11}$
\end{tabular}};

\node[
  box,
  draw=roadmapred,
  fill=roadmapred!10,
  text width=78mm,
  below=3mm of L3
] (L4)
{\textbf{Exact linear-response ceiling}\\[-1pt]
$\displaystyle
H_I^{\max}
=
\frac{2\pi f_I}
{\gamma_a\left|\mathcal T_\theta\mathcal G(\theta_i)\right|}
\sqrt{\mathcal P_{II}^{\max}}$};

\node[
  box,
  draw=roadmapcyan!90!black,
  fill=roadmapcyan!13,
  text width=78mm,
  below=3mm of L4
] (L5)
{\textbf{Minimal all-dark-matter benchmarks}\\[-1pt]
$\theta_i=1,\ f_I=f_a:
\quad ~~~~~~~H_I^{\max}\simeq2.2\times10^7\,\mathrm{GeV}$\\[-1pt]
$\theta_i\ll1,\ f_I=f_a\leq M_{\rm Pl}:
\quad H_I^{\max}\simeq1.25\times10^{10}\,\mathrm{GeV}$};

\foreach \u/\v in {L1/L2,L2/L3,L3/L4,L4/L5}
  \draw[soft] (\u)--(\v);


\node[
  box,
  draw=roadmaporange,
  fill=roadmaporange!7,
  text width=78mm,
  inner ysep=3.8pt,
  below=3mm of RH
] (R1)
{\textbf{Low scale fixes the vacuum energy}\\[-1pt]
$\displaystyle
V_*^{1/4}
\simeq
\left(3M_{\rm Pl}^2H_I^2\right)^{1/4}$};

\node[
  box,
  draw=roadmaporange,
  fill=roadmaporange!7,
  text width=78mm,
  inner ysep=3.8pt,
  below=3mm of R1
] (R2)
{\textbf{Scalar amplitude fixes an extreme slope}\\[-1pt]
$\displaystyle
\epsilon_V
\simeq
\frac{H_I^2}
{8\pi^2M_{\rm Pl}^2A_s}
\sim10^{-16}\text{--}10^{-10}$};

\node[
  box,
  draw=roadmaporange,
  fill=roadmaporange!7,
  text width=78mm,
  inner ysep=4.4pt,
  minimum height=14mm,
  below=3mm of R2
] (R3)
{\textbf{Observed red tilt fixes the curvature}\\[2pt]
$\displaystyle
\eta_V
\simeq
\frac{n_s-1}{2}
\sim-10^{-2}$};

\node[
  box,
  draw=roadmaporange,
  fill=roadmaporange!13,
  text width=78mm,
  inner ysep=4.2pt,
  minimum height=14mm,
  below=3mm of R3
] (R4)
{\textbf{Axion-conditioned derivative hierarchy}\\[-1pt]
$\displaystyle
\left|
M_{\rm Pl}\frac{V'}{V}
\right|
\ll
\left|
M_{\rm Pl}^2\frac{V''}{V}
\right|$};

\node[
  box,
  draw=roadmaporange,
  fill=roadmaporange!9,
  text width=78mm,
  inner ysep=4.2pt,
  minimum height=18mm,
  below=3mm of R4
] (R5)
{\textbf{Axion-conditioned tensor and Lyth bounds}\\[-2pt]
$\displaystyle
\begin{gathered}
r<8\frac{\mathcal P_{II}^{\max}}{A_s}
\left[
\frac{f_I}
{\gamma_a\left|\mathcal T_\theta\mathcal G(\theta_i)\right|M_{\rm Pl}}
\right]^2\\[2pt]
\Longrightarrow\qquad
\frac{\Delta\phi_{\rm CMB}}{M_{\rm Pl}}\ll1
\end{gathered}$};

\node[
  box,
  draw=roadmaporange,
  fill=roadmaporange!7,
  text width=78mm,
  inner ysep=3.8pt,
  below=3mm of R5
] (R6)
{\textbf{Reheating/PQ obstruction}\\[-1pt]
A restricted reheating history lowers $N_*$ and shifts standard plateau and
hilltop predictions toward a redder tilt};

\foreach \u/\v in {R1/R2,R2/R3,R3/R4,R4/R5,R5/R6}
  \draw[soft] (\u)--(\v);

\draw[link]
  (L5.east)--++(4mm,0)|-(R2.west);


\node[
  box,
  draw=roadmaporange,
  fill=roadmaporange!16,
  line width=0.9pt,
  text width=170mm,
  minimum height=14mm,
  anchor=north
] (OBS) at ([yshift=-7mm]tophead.center |- R6.south)
{\textbf{MODEL-BUILDING CRITERION:}
a viable canonical construction must independently control the vacuum energy
$V_0$, the extremely small slope $V'_*$, the percent-level curvature $V''_*$,
the end of inflation, and a reheating history consistent with the assumed PQ
chronology.};


\node[
  head,
  draw=roadmapgreen!90!black,
  fill=roadmapgreen!88!black,
  text=white,
  text width=170mm,
  below=4mm of OBS
] (MH)
{INFLATIONARY MODEL AUDIT (SEC.~5)};

\node[
  modelgood,
  below=3mm of MH,
  xshift=-65.3mm
] (M1)
{\textbf{Pass most directly}\\
Vacuum-dominated hybrid inflation};

\node[
  modelcond,
  right=4mm of M1
] (M2)
{\textbf{Viable with tuning}\\
Running-mass, quadratic hilltop, 
and inflection-point models};

\node[
  modelbad,
  right=4mm of M2
] (M3)
{\textbf{Generally fail a minimal test}\\
Monomials, generic hilltops, Starobinsky and universal plateaus};

\node[
  modelbeyond,
  right=4mm of M3
] (M4)
{\textbf{Modify scalar generation}\\
Warm, noncanonical, curvaton and modulated scenarios};


\node[
  head,
  draw=roadmapviolet!90!black,
  fill=roadmapviolet!92!black,
  text=white,
  text width=170mm,
  below=10mm of $(M2.south)!0.5!(M3.south)$
] (EH)
{WHICH ASSUMPTION OF THE MINIMAL AXION BOUND IS RELAXED? (SEC.~6)};

\node[
  rowlabel,
  below=2mm of EH
] (EIL)
{INFLATIONARY AXION ASSUMPTIONS};

\node[
  escape,
  below=1.5mm of EIL,
  xshift=-56.5mm
] (E11)
{\textbf{Axion light during inflation}\\
Early confinement, small instantons,
temporary explicit PQ breaking,
or topological masses};

\node[
  escape,
  right=5mm of E11
] (E12)
{\textbf{Fixed inflationary radius: $f_I=f_a$}\\
Radial PQ evolution, $f_I/f_a\gg1$, curved field space, or nonminimal
gravitational couplings};

\node[
  escape,
  right=5mm of E12
] (E13)
{\textbf{Fixed minimum and alignment}\\
Moving minima, discrete gauge symmetries, sequestering, alignment, or
topological mass generation};

\node[
  rowlabel,
  below=2mm of E12
] (EPL)
{POST-INFLATIONARY AXION ASSUMPTIONS};

\node[
  escape,
  below=1.5mm of EPL,
  xshift=-56.5mm
] (E21)
{\textbf{PQ broken throughout the history}\\
Post-inflationary restoration or breaking, followed by strings, walls, or
other defect dynamics};

\node[
  escape,
  right=5mm of E21
] (E22)
{\textbf{Standard relic history}\\
Kinetic misalignment, Witten effect damping,
resonance or fragmentation, entropy dilution,
and modified thermal evolution};

\node[
  escape,
  right=5mm of E22
] (E23)
{\textbf{Standard post-exit conversion}\\
Axion mixing, level crossing, entropy transfer, curvaton effects, or
modulated reheating};


\node[
  box,
  draw=roadmapblue,
  fill=roadmapblue,
  text=white,
  line width=0.9pt,
  text width=170mm,
  minimum height=13mm,
  below=5mm of E22
] (TAKE)
{\textbf{TAKE-HOME MESSAGE:}
either realize the axion-conditioned hierarchy in the inflationary sector or
relax a specific assumption of the minimal axion isocurvature bound. Any
escape mechanism must also control the new dynamics, tuning, defects,
reheating effects, or ultraviolet sensitivity that it introduces.};


\begin{scope}[on background layer]

\node[
  rounded corners=2pt,
  draw=roadmapred!72!black,
  fill=roadmapred!4,
  line width=0.5pt,
  fit=(LH)(L1)(L2)(L3)(L4)(L5),
  inner sep=3mm
] {};

\node[
  rounded corners=2pt,
  draw=roadmaporange!74!black,
  fill=roadmaporange!4,
  line width=0.5pt,
  fit=(RH)(R1)(R2)(R3)(R4)(R5)(R6),
  inner sep=3mm
] {};

\node[
  rounded corners=2pt,
  draw=roadmapgreen!72!black,
  fill=roadmapgreen!4,
  line width=0.5pt,
  fit=(MH)(M1)(M2)(M3)(M4),
  inner sep=3mm
] (MPANEL) {};

\node[
  rounded corners=2pt,
  draw=roadmapviolet!72!black,
  fill=roadmapviolet!4,
  line width=0.5pt,
  fit=(EH)(EIL)(E11)(E12)(E13)(EPL)(E21)(E22)(E23),
  inner sep=3mm
] (EPANEL) {};

\end{scope}

\end{tikzpicture}
\end{adjustbox}

\caption{Logical structure of the analysis. The minimal pre-inflationary
axion history generates the isocurvature constraint shown at the top and
left. Its implications for the inflationary scale, potential derivatives,
field excursion, and reheating history motivate the inflationary model audit
in Sec.~\ref{sec:models}. The complementary possibility is to relax a
specific assumption of the minimal axion bound through modified axion or PQ
dynamics, as classified in Sec.~\ref{sec:loopholes}.}
\label{fig:roadmap}
\end{figure}

This paper develops these observations into a quantitative model-building
filter. We first derive the exact flat-spectrum isocurvature bound for a
light axion, keeping the inflationary canonical radius $\fI$ distinct from
the late-time QCD axion decay constant $\fa$, allowing an arbitrary axion
dark-matter fraction, and retaining the full anharmonic response of the
misalignment abundance. We then evaluate the bound using current Planck,
ACT, and SPT constraints and identify the largest inflationary scale allowed
within the minimal scenario. The maximum occurs only near
$\fa\simeq\Mpl$ with a very small initial angle. An order-one angle gives a
ceiling of order $10^7\GeV$, while near-hilltop initial conditions can
strengthen the limit by many additional orders of magnitude.

We next combine the isocurvature and tensor spectra to derive a direct
axion--tensor inequality and an axion-conditioned counterpart of the Lyth
bound. We show how the resulting restriction on the local inflaton slope,
together with the measured tilt and the reheating history, selects models in
which the vacuum energy, slope, curvature, end of inflation, and PQ sector
can be controlled with substantial independence. Vacuum-dominated hybrid
inflation realizes this separation most directly, while running-mass and
inflection-point constructions provide viable but generally more
ultraviolet-sensitive alternatives. By contrast, a simple rescaling of a
successful high-scale potential is usually insufficient.

The minimal result is not a no-go theorem for QCD axion dark matter. Its
assumptions can be relaxed by changing the axion mass during inflation, its
canonical normalization, the PQ chronology, the relic abundance, or the
post-exit conversion of axion fluctuations into the late-time isocurvature
mode. We therefore organize nonminimal proposals according to the assumption
of the minimal bound that they modify. This separates mechanisms that
suppress the primordial fluctuation from those that alter its transfer or
dilute its observational effect, and makes clear which dynamical or
ultraviolet requirements replace the original constraint.

Fig.~\ref{fig:roadmap} summarizes the structure of the analysis. The
minimal pre-inflationary history produces the isocurvature bound shown in the
upper part of the figure. Its consequences for the inflationary potential,
field excursion, and reheating history lead to the model audit in
Sec.~\ref{sec:models}, while modifications of the axion sector are classified
in Sec.~\ref{sec:loopholes}.

The paper is organized as follows. Sec.~\ref{sec:axionbound} states the
assumptions of the minimal scenario, derives the exact isocurvature bound,
and evaluates it using current CMB limits. Sec.~\ref{sec:lyth} develops
the axion--tensor relation, the axion-conditioned Lyth bound, and the
resulting hierarchy of potential derivatives.
Sec.~\ref{sec:reheating} incorporates the reheating chronology and its
implications for the scalar tilt and PQ non-restoration.
Sec.~\ref{sec:models} assesses inflationary model classes against these
requirements. Sec.~\ref{sec:loopholes} classifies the axion
mechanisms that suppress or evade the minimal bound. We conclude in
Sec.~\ref{sec:conclusions}.

\section{The pre-inflationary axion bound}
\label{sec:axionbound}
\subsection{Assumptions and the distinction between
\texorpdfstring{$f_I$}{fI} and \texorpdfstring{$f_a$}{fa}}
\label{subsec:assumptions}
The standard pre-inflationary axion isocurvature bound rests on a specific set
of assumptions, which we state explicitly. We consider a history in which the
PQ symmetry is broken before the largest observable CMB scale exits the
horizon and is never subsequently restored, either thermally or
nonthermally. The observable Universe then descends from a single PQ domain,
and the axion is present during inflation as a weakly coupled spectator
field~\cite{Axenides:1983hj,Seckel:1985tj,Linde:1985yf,Lyth:1989pb,
Turner:1990uz,Lyth:1991ub,Lyth:1992tw}. At the pivot scale, we assume
$m_{a,I}^2\ll \HI^2$, where $m_{a,I}$ is the effective mass of the axion
fluctuation on the inflationary background, and that its canonical radius
varies slowly across the observable window. Together with negligible mixing
with the inflaton, these conditions give a nearly scale-invariant
horizon-exit spectrum, with corrections controlled by
$m_{a,I}^2/\HI^2$, the inflationary slow-roll parameters, and the variation
of the canonical radius. Any subsequent superhorizon evolution is encoded in
the transfer factor introduced below.

We further assume that the axion perturbation is approximately Gaussian and
statistically uncorrelated with the primordial curvature perturbation. After
inflation, the axion abundance is generated by standard vacuum
misalignment~\cite{Preskill:1982cy,Abbott:1982af,Dine:1982ah}, without
substantial late entropy production or conversion between the axion
isocurvature and adiabatic modes. Finally, we work in the linear response
regime, expanding the relic abundance about the homogeneous initial
misalignment angle. These assumptions define the baseline one field
scenario. We keep the inflationary normalization and perturbation transfer
explicit in the general bound. The minimal numerical benchmark further sets
$\fI=\fa$ and $\Ttheta=1$. Sec.~\ref{sec:loopholes} discusses mechanisms
that relax one or more of these assumptions.

It is essential to distinguish the canonical radius of the axion direction
at horizon exit, denoted by $\fI$, from the late-time QCD axion decay constant
$\fa$. Let $\theta_*$ be the dimensionless angular coordinate when the pivot
mode exits the horizon. For an effectively massless canonical angular mode,
the dimensionless horizon-exit power-spectrum amplitude is
\begin{equation}
  \delta\theta_*
  =
  \frac{\HI}{2\pi\fI} \, .
  \label{eq:deltatheta}
\end{equation}
Here $\fI$ is the field-space radius of the angular direction that later
evolves into the QCD axion. It is therefore $\fI$, rather than $\fa$, that
sets the primordial angular variance.\footnote{In simple single-field PQ
models, $\fI$ coincides with the quantity often denoted
$f_a^{(\mathrm{inf})}$. A hierarchy $\fI\gg\fa$ can suppress the
horizon-exit angular variance, but requires the PQ background or its kinetic
normalization to differ during inflation. The subsequent evolution must also
avoid excessive axion production, parametric resonance, and nonthermal PQ
restoration~\cite{Linde:1991km,Nakayama:2015pba,Harigaya:2015hha,
Kobayashi:2016qld,Graham:2025iwx}.}

In the minimal evolution assumed below, both the homogeneous angular
coordinate and its superhorizon perturbation are conserved between horizon
exit and the onset of the QCD potential,
\begin{equation}
  \thetai=\theta_*,
  \qquad
  \delta\thetai=\delta\theta_*.
  \label{eq:angletransferminimal}
\end{equation}
For a single effective angular degree of freedom, a more general mapping may
be written as
\begin{equation}
  \thetai=\thetai(\theta_*),
  \qquad
  \delta\thetai
  =
  \Ttheta\,\delta\theta_*,
  \qquad
  \Ttheta
  \equiv
  \frac{\partial\thetai}{\partial\theta_*}.
  \label{eq:angletransfergeneral}
\end{equation}
A nontrivial scalar transfer factor can arise from nonadiabatic evolution of
the PQ sector~\cite{Kobayashi:2016qld}. With several light angular fields,
the corresponding relation is generally matrix-valued.
Eq.~\eqref{eq:angletransfergeneral} is then the one-dimensional reduction
relevant to the late QCD axion direction. We retain $\Ttheta$ in the general
one field bound and set $\Ttheta=1$ in the minimal numerical benchmark.

The transfer function $\Ttheta$ is conceptually distinct from the hierarchy
$\fI/\fa$. Although a given model can correlate them, they enter the bound in
different ways: $\fI$ fixes the horizon-exit fluctuation, whereas $\Ttheta$
describes its subsequent linear evolution. An evolving PQ radius or a
nonminimal gravitational coupling can yield $\fI\neq\fa$ even when the
angular coordinate is conserved~\cite{Graham:2025iwx,Rigouzzo:2025ycb}.
Conversely, curved field-space dynamics can generate an effective angular
mass or coupled entropy evolution during inflation. In that case, the
horizon-exit spectrum and its superhorizon transfer must be computed
together, rather than modifying Eq.~\eqref{eq:deltatheta} by a post-horizon
factor alone. We discuss these possibilities in Sec.~\ref{sec:loopholes}.
Substituting $\fa$ into Eq.~\eqref{eq:deltatheta} without first determining
the inflationary canonical normalization can therefore lead to
parametrically incorrect conclusions.

\subsection{The isocurvature spectrum and the bound on
\texorpdfstring{$H_I$}{HI}}
\label{subsec:power}
We define the fraction of the present-day cold dark matter abundance carried
by axions as
\begin{equation}
  \gamma_a
  \equiv
  \frac{\Omegaa}{\OmegaDM},
  \label{eq:gamma}
\end{equation}
where $\OmegaDM$ is the present-day cold dark matter density parameter. At fixed $\fa$ and fixed background cosmological parameters, we define the
logarithmic response of the axion relic abundance to the initial
misalignment angle by
\begin{equation}
  \left.
  \frac{\delta\Omegaa}{\Omegaa}
  \right|_{\delta\rho_\gamma=0}
  =
  \Gresp(\thetai)\,\delta\thetai,
  \qquad
  \Gresp(\thetai)
  \equiv
  \left.
  \frac{\partial\ln\Omegaa}{\partial\thetai}
  \right .
  \label{eq:Gdefinition}
\end{equation}
Here $\delta\Omegaa$ denotes the variation of the late-time relic abundance
between neighboring separate-universe patches, rather than a perturbation of
the local critical density. The response $\Gresp$ includes the full dependence of the relic abundance on the initial angle, in particular the anharmonic evolution that becomes important near the hilltop~\cite{Lyth:1991ub,Visinelli:2009zm,
Kobayashi:2013nva}.

After coherent oscillations begin and the axion behaves as pressureless
matter, its gauge-invariant entropy perturbation relative to radiation is
\begin{equation}
  S_a
  \equiv
  3\left(\zeta_a-\zeta_\gamma\right)
  =
  \frac{\delta\rho_a}{\rho_a}
  -
  \frac{3}{4}
  \frac{\delta\rho_\gamma}{\rho_\gamma},
  \label{eq:SaGI}
\end{equation}
where $\zeta_a$ and $\zeta_\gamma$ are the curvature perturbations on
uniform axion-density and uniform radiation-density hypersurfaces,
respectively. Comparing separate superhorizon patches on a uniform-radiation
hypersurface at a common post-oscillation epoch gives
\begin{equation}
  S_a
  =
  \left.
  \frac{\delta\rho_a}{\rho_a}
  \right|_{\delta\rho_\gamma=0}
  =
  \left.
  \frac{\delta\Omegaa}{\Omegaa}
  \right|_{\delta\rho_\gamma=0}
  =
  \Gresp(\thetai)\,\delta\thetai.
  \label{eq:Saresponse}
\end{equation}
Here the abundance is evaluated at fixed $\fa$ and fixed
background cosmology, as in Eq.~\eqref{eq:Gdefinition}.

If the non-axion cold dark matter carries only the adiabatic perturbation,
then, at linear order, the total cold dark matter entropy perturbation is $S_{\rm cdm}=\gamma_a S_a$, once both components behave as pressureless matter~\cite{Fox:2004kb, Beltran:2006sq}.

Following the Planck convention, we denote the primordial curvature and cold
dark matter isocurvature (CDI) power spectra by $\PRR$ and $\PII$, and their
cross-spectrum by $\mathcal P_{\mathcal RI}$~\cite{Planck:2018jri}. We
consider the uncorrelated case, $\mathcal P_{\mathcal RI}=0$, and identify
the CDI perturbation with the total cold dark matter entropy perturbation,
$\mathcal I=S_{\rm cdm}$.

For a mode of wavenumber $k$, the transfer factor in
Eq.~\eqref{eq:angletransfergeneral} may in general depend on its horizon-exit
time. The CDI perturbation is therefore
\begin{equation}
  S_{\rm cdm}(k)
  =
  \gamma_a\Gresp(\thetai)\Ttheta(k)\delta\theta_*(k),
  \label{eq:Scdmtransfer}
\end{equation}
and its power spectrum is
\begin{equation}
  \PII(k)
  =
  \left[
    \gamma_a\Gresp(\thetai)\Ttheta(k)
  \right]^2
  \mathcal P_{\delta\theta_*}(k).
  \label{eq:PIIfromtheta}
\end{equation}
At the pivot scale we henceforth write
$\Ttheta\equiv\Ttheta(k_*)$. Using
$\mathcal P_{\delta\theta_*}(k_*)=(\HI/2\pi\fI)^2$ then gives
\begin{equation}
  \PII(k_*)
  =
  \left[
    \gamma_a\,
    \Gresp(\thetai)\,
    \Ttheta\,
    \frac{\HI}{2\pi\fI}
  \right]^2.
  \label{eq:PIIexact}
\end{equation}

We parameterize the resulting primordial CDI spectrum as a power law,
\begin{equation}
  \PII(k)
  =
  \PII(k_*)
  \left(
    \frac{k}{k_*}
  \right)^{n_{II}-1},
  \label{eq:PIItilt}
\end{equation}
where $n_{II}=1$ is scale invariant, while $n_{II}<1$ and $n_{II}>1$
correspond to red and blue spectra, respectively. Any sufficiently mild
scale dependence of the horizon-exit spectrum or of $\Ttheta(k)$ is included
in $n_{II}$. A strongly non-power-law transfer requires the full spectrum
rather than Eq.~\eqref{eq:PIItilt}.

For an uncorrelated CDI mode, the isocurvature fraction at the pivot scale is
\begin{equation}
  \betaiso(k_*)
  \equiv
  \frac{\PII(k_*)}
       {\PRR(k_*)+\PII(k_*)}.
  \label{eq:betaexact}
\end{equation}
At $k_*=0.05\,\mathrm{Mpc}^{-1}$, $\PRR(k_*)=\As$, with
$\As\simeq2.10\times10^{-9}$ for the Planck 2018 analysis including TT, TE,
EE, low-$\ell$ polarization, and lensing~\cite{Planck:2018vyg}. At any fixed point in
parameter space, Eq.~\eqref{eq:betaexact} implies the exact identity
\begin{equation}
  \PII(k_*)
  =
  \As
  \frac{\betaiso(k_*)}
       {1-\betaiso(k_*)}.
  \label{eq:PIIbetaidentity}
\end{equation}

Combining Eqs.~\eqref{eq:PIIexact} and \eqref{eq:PIIbetaidentity} yields
\begin{equation}
  \HI
  <
  \frac{2\pi\fI}
       {\gamma_a\,|\Ttheta\Gresp(\thetai)|}
  \left[
    \As
    \frac{\betaiso^{\rm max}(k_*)}
         {1-\betaiso^{\rm max}(k_*)}
  \right]^{1/2}.
  \label{eq:masterboundbeta}
\end{equation}
If the likelihood instead provides a direct upper limit on the CDI amplitude,
the bound is
\begin{equation}
  \HI
  <
  \frac{2\pi\fI}
       {\gamma_a\,|\Ttheta\Gresp(\thetai)|}
  \sqrt{\PII^{\rm max}(k_*)}.
  \label{eq:masterboundPII}
\end{equation}
Equations~\eqref{eq:masterboundbeta} and \eqref{eq:masterboundPII} are
algebraically equivalent when $\betaiso$, $\PII$, and $\As$ are evaluated at
the same point in parameter space. Their separately marginalized confidence
limits cannot, however, generally be converted into one another by inserting
a single central value of $\As$, because these parameters are correlated in
the likelihood. We therefore use the directly reported limits on $\PII$ from
Ref.~\cite{Petretti:2026ayw} for the scale-invariant numerical results below.

For minimal transfer, $\Ttheta=1$, the harmonic regime gives
\begin{equation}
  \Omegaa\propto\thetai^2,
  \qquad
  \Gresp(\thetai)=\frac{2}{\thetai},
\end{equation}
and Eq.~\eqref{eq:masterboundbeta} reduces to
\begin{equation}
  \HI
  <
  \frac{\pi\fI|\thetai|}{\gamma_a}
  \left[
    \As
    \frac{\betaiso^{\rm max}(k_*)}
         {1-\betaiso^{\rm max}(k_*)}
  \right]^{1/2}.
  \label{eq:harmonicbound}
\end{equation}
At fixed $\thetai$, this bound scales as $\gamma_a^{-1}$. If the abundance
condition is imposed, however, $\thetai$ is no longer independent. In the
small-angle standard-misalignment regime, at fixed $\fa$ and fixed background
cosmology, $\Omegaa\propto\thetai^2$ and hence
$|\thetai|\propto\sqrt{\gamma_a}$. Consequently, when $\fI=\fa$, the
abundance-conditioned bound scales as
$\HI^{\rm max}\propto\gamma_a^{-1/2}$ rather than $\gamma_a^{-1}$.

\subsection{Benchmark abundance and current CMB limits}
\label{subsec:benchmark}
For numerical comparisons, we use the benchmark QCD axion abundance fit~\cite{Lyth:1991ub,Strobl:1994wk,Bae:2008ue,Visinelli:2009zm,
Dine:2017swf,Petretti:2026ayw}
\begin{align}
  \Omegaa h^2
  &\simeq
  0.12\,\widetilde\theta_i^{\,2}
  \left(
    \frac{\fa}{10^{12}\GeV}
  \right)^{7/6},
  \label{eq:abundancefit}
  \\
  \widetilde\theta_i^{\,2}
  &\equiv
  \thetai^2
  \left[
    \ln\!\left(
      \frac{e}{1-\thetai^2/\pi^2}
    \right)
  \right]^{7/6}.
  \label{eq:thetatile}
\end{align}
The logarithmic factor parametrizes the anharmonic enhancement of the relic
abundance as $|\thetai|\to\pi$. For $h\simeq0.67$, the normalization in
Eq.~\eqref{eq:abundancefit} is equivalent to $\Omegaa\simeq0.27$, the form
used in Ref.~\cite{Petretti:2026ayw}.

Eqs.~\eqref{eq:abundancefit} and \eqref{eq:thetatile} are a convenient
benchmark, not a substitute for evolving the axion with a temperature-dependent QCD susceptibility~\cite{GrillidiCortona:2015jxo, Borsanyi:2016ksw}. In particular, at fixed $\thetai$, the scaling
$\Omegaa\propto\fa^{7/6}$ assumes that oscillations begin while the axion
mass remains appreciably temperature dependent. At sufficiently large
$\fa$, their onset moves into the crossover toward the low-temperature,
approximately constant mass regime, where the asymptotic scaling approaches
$\Omegaa\propto\fa^{3/2}$ rather than $\fa^{7/6}$~\cite{Visinelli:2009zm,
Marsh:2015xka}. Numerical extrema obtained by extrapolating
Eq.~\eqref{eq:abundancefit} to $\fa\sim\Mpl$ are therefore conditional on
this benchmark extrapolation and should not be interpreted as precision QCD
predictions. We retain the fit throughout to provide a transparent common
benchmark and to compare directly with Ref.~\cite{Petretti:2026ayw}.

The isocurvature bound depends on the logarithmic response of the abundance,
rather than on its overall normalization. The normalization enters only when
the abundance condition relating $\gamma_a$, $\thetai$, and $\fa$ is
imposed. Defining
\begin{equation}
  L(\thetai)
  \equiv
  \ln\!\left(
    \frac{e}{1-\thetai^2/\pi^2}
  \right),
  \label{eq:Ldef}
\end{equation}
the response associated with
Eqs.~\eqref{eq:abundancefit}--\eqref{eq:thetatile} is
\begin{equation}
  \Gresp(\thetai)
  =
  \frac{2}{\thetai}
  +
  \frac{7\thetai}
       {3\pi^2
       \left(1-\thetai^2/\pi^2\right)
       L(\thetai)}.
  \label{eq:Ganalytic}
\end{equation}
For definiteness, we take $0<\thetai<\pi$. The response diverges in both the
harmonic limit, $\thetai\to0$, and the hilltop limit,
$\thetai\to\pi$. For the adopted fit, it has a unique minimum at
$\thetai\simeq2.124$, where
$\Gresp^{\rm min}\simeq1.516$. This corresponds to
$\thetai\simeq121.7^\circ$, and the anharmonic term contributes about
$38\%$ of the total response.

The linear-response treatment also requires
$|\Gresp(\thetai)|\,\delta\thetai\ll1$. Near the hilltop, the stronger
condition $\delta\thetai\ll\pi-\thetai$ ensures that the fluctuation does not
sample the nonlinear boundary of the angular potential. If either condition
fails, a stochastic and generally non-Gaussian calculation replaces the
Gaussian power-spectrum bound used here.

At fixed $\fI$, $\gamma_a$, and observational limit on $\PII$, the angular
dependence of the bound is controlled entirely by
$1/|\Gresp(\thetai)|$. The constraint is therefore least restrictive near
the minimum of $\Gresp$ and becomes stronger toward both the harmonic and
hilltop limits. This is a local statement in parameter space and should not
be confused with maximizing the allowed $\HI$ along an
abundance-compatible trajectory, on which $\fa$, and possibly $\fI$, also
vary.

The hilltop regime illustrates the distinction. Anharmonic enhancement
allows a fixed relic abundance to be obtained at a smaller $\fa$, but it
also increases the sensitivity of the abundance to the initial angle. The
divergence of $\Gresp$ as $\thetai\to\pi$ therefore strengthens the
isocurvature constraint rather than providing a hilltop loophole within the
linear regime~\cite{Kobayashi:2013nva}.

We now specify the observational limits used below. Ref.~\cite{Petretti:2026ayw}
analyses the Planck PR4/NPIPE \textsc{CamSpec} high-$\ell$ likelihood~\cite{Rosenberg:2022sdy} alone and in combination with ACT DR6~\cite{AtacamaCosmologyTelescope:2025blo} or SPT-3G D1~\cite{SPT-3G:2025bzu}. We denote the two combinations by P-ACT and P-SPT.
For an uncorrelated, scale-invariant CDI spectrum,
$n_{II}=1$, the marginalized $95\%$ upper limits at
$k_*=0.05\,\mathrm{Mpc}^{-1}$ are shown in
Table~\ref{tab:CMBlines}.

\begin{table}[t]
\centering
\begin{tabular}{lcc}
\toprule
Data set
&
$\PII^{\rm max}(k_*)$
&
$\betaiso^{\rm max}(k_*)$
\\
\midrule
Planck
&
$5.30\times10^{-11}$
&
$0.0245$
\\
P-ACT
&
$7.51\times10^{-11}$
&
$0.0336$
\\
P-SPT
&
$5.72\times10^{-11}$
&
$0.0262$
\\
\bottomrule
\end{tabular}
\caption{Marginalized $95\%$ upper limits on an uncorrelated,
scale-invariant CDI mode at $k_*=0.05\,\mathrm{Mpc}^{-1}$. The limits on
$\PII$ and $\betaiso$ are reported separately in
Ref.~\cite{Petretti:2026ayw}. Their posterior endpoints therefore need not be
related by inserting a single fixed value of $\As$.}
\label{tab:CMBlines}
\end{table}

The Planck result in Table~\ref{tab:CMBlines} uses the PR4/NPIPE
\textsc{CamSpec} high-$\ell$ likelihood~\cite{Rosenberg:2022sdy}. Using the
\texttt{plik\_lite} likelihood instead gives the weaker limit
$\PII^{\rm max}\simeq9.34\times10^{-11}$~\cite{Petretti:2026ayw}. The difference between the bound quoted here and
looser Planck-only limits appearing in some earlier analyses is therefore
driven substantially by the high-$\ell$ likelihood choice, not only by the
addition or removal of a particular experiment.

A notable feature of the fixed, scale-invariant fits is that adding ACT or
SPT does not strengthen the marginalized CDI-amplitude constraint.
P-ACT gives the weakest of the three limits, P-SPT is intermediate,
and Planck alone gives the strongest. This behavior reflects two related
effects. First, for flat and moderately blue primordial spectra, the CDI
transfer functions fall rapidly at the high multipoles where ACT and SPT have
their greatest statistical power, limiting the direct gain from additional
small-scale measurements. Second, the CDI amplitude is correlated with the
adiabatic parameters, especially the scalar tilt $n_s$. For fixed
$n_{II}\geq1$, a value of $n_s$ closer to unity lowers the adiabatic power on
large scales relative to a redder spectrum and can therefore allow a larger
CDI contribution. Since P-ACT favors the largest $n_s$ among the three
combinations, it also permits the largest marginalized CDI amplitude~\cite{Petretti:2026ayw}.

This ordering is not universal across spectral shapes. For the strongly red
benchmark $n_{II}=0$, Ref.~\cite{Petretti:2026ayw} finds that P-ACT and
P-SPT are slightly more constraining than Planck alone. We return to the
case of a varying isocurvature tilt in Sec.~\ref{sec:loopholes}.

\subsection{The largest allowed inflationary scale}
\label{subsec:Hmax}

We now evaluate the minimal benchmark defined by
\begin{equation}
  \fI=\fa,
  \qquad
  \Ttheta=1,
  \qquad
  \Omegaa=\OmegaDM.
  \label{eq:minimalbenchmark}
\end{equation}
For each value of $\fa$, the abundance condition determines the initial
misalignment angle $\thetai(\fa)$ through
Eqs.~\eqref{eq:abundancefit}--\eqref{eq:thetatile}. The corresponding
$95\%$ C.L. upper limit on the inflationary Hubble scale is therefore the
one-dimensional function
\begin{equation}
  \HI^{\rm max}(\fa)
  =
  \frac{2\pi\fa}
       {\left|\Gresp[\thetai(\fa)]\right|}
  \sqrt{\PII^{\rm max}},
  \label{eq:Hmaxfaexact}
\end{equation}
where $\PII^{\rm max}\equiv\PII^{\rm max}(k_*)$ and
$0<\thetai(\fa)<\pi$ is the solution of
\begin{equation}
  \widetilde\theta_i^{\,2}(\thetai)
  \left(
    \frac{\fa}{10^{12}\GeV}
  \right)^{7/6}
  =
  \frac{\OmegaDM h^2}{0.12}.
  \label{eq:abundanceconditionfa}
\end{equation}
For the benchmark value $\OmegaDM h^2\simeq0.12$ used in
Eq.~\eqref{eq:abundancefit}, the right-hand side is approximately unity.

\begin{figure}[t]
\centering
\IfFileExists{figures/axion_isocurvature_envelope.pdf}{%
  \includegraphics[width=0.91\textwidth]
  {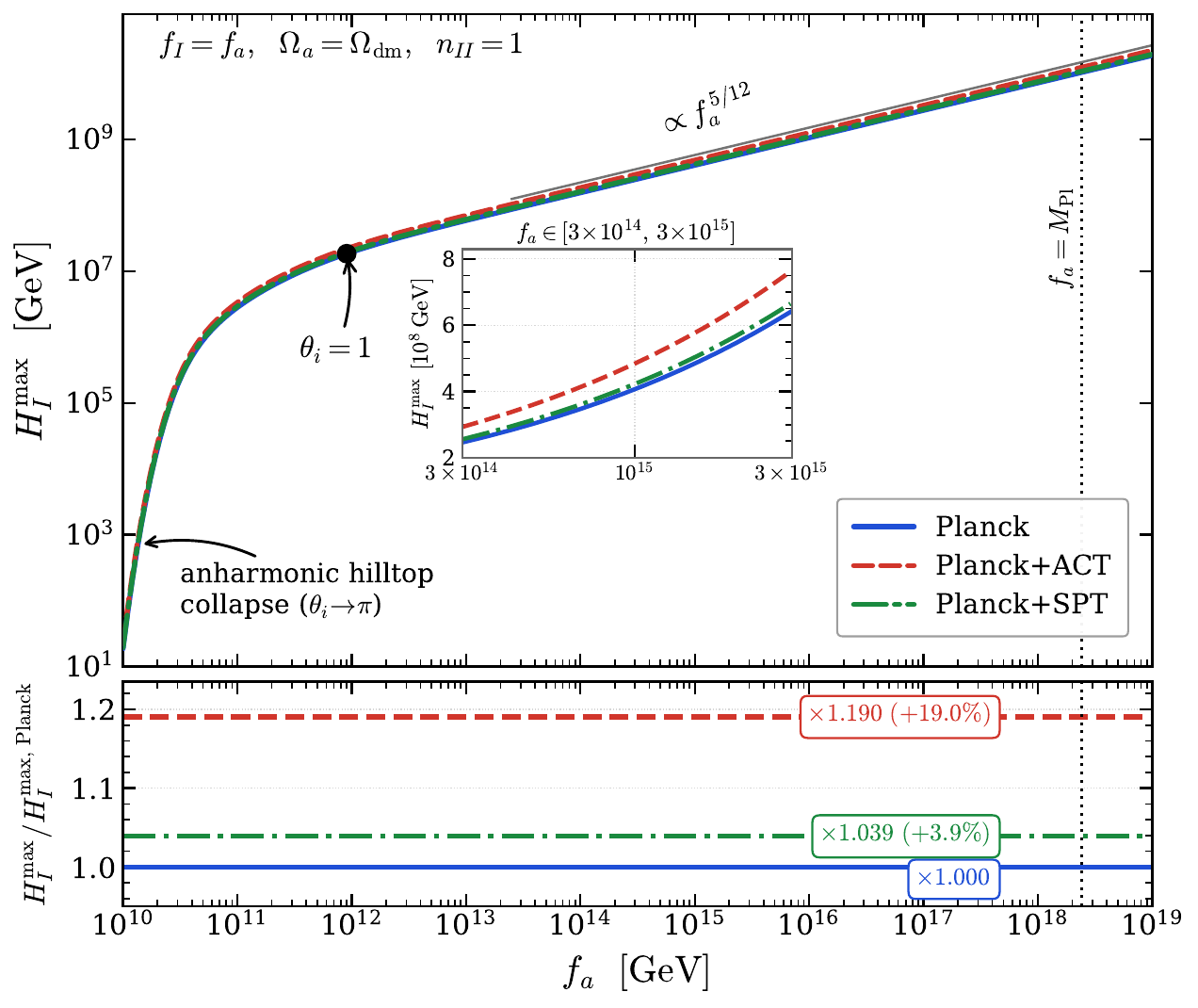}
}{%
  \fbox{%
    \parbox{0.85\textwidth}{%
      \centering
      Missing file
      \texttt{axion\_isocurvature\_envelope.pdf}.
    }%
  }
}
\caption{Maximum allowed inflationary Hubble scale as a function of the QCD
axion decay constant in the minimal pre-inflationary scenario, using the
benchmark abundance fit in Eq.~\eqref{eq:abundancefit} with $\fI=\fa$,
$\Ttheta=1$, and $\Omegaa=\OmegaDM$. At large $\fa$, the abundance
condition requires a small initial angle and, within the adopted
$\fa^{7/6}$ fit, the curves approach
$\HI^{\rm max}\propto\fa^{5/12}$. At smaller $\fa$, the required angle
approaches the hilltop, where the rapidly increasing anharmonic response
strongly tightens the isocurvature bound.}
\label{fig:axionenvelope}
\end{figure}

Fig.~\ref{fig:axionenvelope} shows Eq.~\eqref{eq:Hmaxfaexact} for the
observational limits in Table~\ref{tab:CMBlines}. The largest value within
the benchmark is found by maximizing $\HI^{\rm max}(\fa)$ over $\fa$,
subject to an explicit upper prior on the decay constant.

The bound exhibits two qualitative regimes, separated roughly near
$\thetai\simeq1$, corresponding to
$\fa\simeq9.03\times10^{11}\GeV$ for the adopted fit. At larger $\fa$, the
required angle is small, the response is approximately harmonic, and
$\HI^{\rm max}(\fa)$ rises slowly with $\fa$. At smaller $\fa$, the
abundance condition drives the angle toward the hilltop, the anharmonic
contribution in Eq.~\eqref{eq:Ganalytic} becomes increasingly important, and
the maximum allowed $\HI$ falls rapidly.

The choice $\thetai=1$ is not physically preferred and does not maximize the
allowed inflationary scale. It is simply a representative order-one
misalignment angle that lies neither in the small-angle regime nor close to
the hilltop. For the adopted abundance fit,
\begin{equation}
  \thetai=1
  \qquad\Longleftrightarrow\qquad
  \fa\simeq9.03\times10^{11}\GeV.
  \label{eq:thetaonefa}
\end{equation}

\begin{table}[t]
\centering
\begin{tabular}{lccc}
\toprule
Data set
&
$\thetai=1$
&
$\fa=10^{15}\GeV$
&
$\fa=\Mpl$
\\
\midrule
Planck
&
$1.85\times10^7\GeV$
&
$4.07\times10^8\GeV$
&
$1.05\times10^{10}\GeV$
\\
P-ACT
&
$2.20\times10^7\GeV$
&
$4.84\times10^8\GeV$
&
$1.25\times10^{10}\GeV$
\\
P-SPT
&
$1.92\times10^7\GeV$
&
$4.22\times10^8\GeV$
&
$1.09\times10^{10}\GeV$
\\
\bottomrule
\end{tabular}
\caption{Benchmark $95\%$ C.L. upper limits on $\HI$ obtained from the full
angular response of the adopted abundance fit in the minimal scenario. The
first column uses the representative order-one angle $\thetai=1$,
corresponding to $\fa=9.03\times10^{11}\GeV$. The remaining columns impose
the all-dark-matter abundance condition, $\Omegaa=\OmegaDM$, at the stated
values of $\fa$. Here $\Mpl=2.435\times10^{18}\GeV$ is the reduced Planck
mass.}
\label{tab:Hmax}
\end{table}

The largest value in Table~\ref{tab:Hmax} is
\begin{equation}
  \HI < 1.25\times10^{10}\GeV
  \qquad (95\%~\mathrm{C.L.}),
\end{equation}
obtained from the P-ACT amplitude limit after imposing the prior
$\fa\leq\Mpl$. This is a prior-dependent ceiling within the adopted
benchmark, rather than a model-independent QCD prediction. It assumes
$\fI=\fa$, a light axion during inflation, negligible classical evolution
of the axion background, $\Ttheta=1$, and $\Omegaa=\OmegaDM$. It also
extrapolates the benchmark $\fa^{7/6}$ abundance law through the crossover
to the low-temperature, approximately constant-mass regime, where that
scaling is no longer accurate.

Relaxing any of these assumptions can shift the numerical ceiling. Examples
include $\fI>\fa$, a non-negligible axion mass during inflation, a
subdominant axion fraction, a nontrivial transfer factor, or an ultraviolet
completion that permits $\fa>\Mpl$. Conversely, taking $\fa$ close to
$\Mpl$ makes it increasingly difficult to protect the required approximate
PQ symmetry from Planck-suppressed symmetry-violating operators and other
quantum gravity effects~\cite{Kamionkowski:1992mf,Banks:2010zn}. The
appropriate order-of-magnitude summary within this benchmark is therefore
$\HI\lesssim10^{10}\GeV$ in the most permissive corner of the minimal
pre-inflationary scenario~\cite{Hertzberg:2008wr,Marsh:2015xka}.

For the representative order-one choice $\thetai=1$, the ceiling is
considerably lower,
\begin{equation}
  \HI < 2.2\times10^7 \, \GeV
  \qquad (95\%~\mathrm{C.L.}) \, ,
\end{equation}
using the least restrictive of the three fixed-$n_{II}=1$ amplitude limits
in Table~\ref{tab:CMBlines}. This value is only a useful reference point.
Moving toward the hilltop tightens the bound because the abundance becomes
increasingly sensitive to the initial angle, whereas moving to smaller
angles requires a larger $\fa$ to maintain $\Omegaa=\OmegaDM$.

The $\mathcal O(10^{10}\GeV)$ ceiling is reached only as $\fa$ approaches
the imposed reduced Planck scale prior. At $\fa=\Mpl$, the adopted abundance
fit gives $\thetai\simeq1.88\times10^{-4}$. Although
$\Gresp\simeq2/\thetai$ grows as the angle decreases, the explicit increase
of $\fI=\fa$ dominates, yielding the slow net scaling derived below.

In the harmonic large $\fa$ regime, $L(\thetai)\simeq1$ and $\widetilde\theta_i\simeq\thetai$. The
$\Omegaa=\OmegaDM$ abundance condition therefore gives
\begin{equation}
  \thetai(\fa)
  \simeq
  \left(
    \frac{\OmegaDM h^2}{0.12}
  \right)^{1/2}
  \left(
    \frac{\fa}{10^{12}\GeV}
  \right)^{-7/12}.
  \label{eq:thetascalinggeneral}
\end{equation}
Combining this result with $\Gresp=2/\thetai$ and $\fI=\fa$ gives
\begin{align}
  \HI^{\rm max}(\fa)
  &\simeq
  \pi\fa\,\thetai(\fa)\sqrt{\PII^{\rm max}}
  \nonumber\\
  &\simeq
  2.72\times10^7\GeV
  \left(
    \frac{\OmegaDM h^2}{0.12}
  \right)^{1/2}
  \left(
    \frac{\PII^{\rm max}}
         {7.51\times10^{-11}}
  \right)^{1/2}
  \left(
    \frac{\fa}{10^{12}\GeV}
  \right)^{5/12}.
  \label{eq:scalingH}
\end{align}
Equation~\eqref{eq:scalingH} is the small-angle asymptote of the benchmark
function in Eq.~\eqref{eq:Hmaxfaexact}. It accurately reproduces the full
benchmark result at large $\fa$ and overestimates it by approximately
$17\%$ at $\fa=10^{12}\GeV$, where the exact response
$\Gresp(\thetai)$ departs from its harmonic form $2/\thetai$.

The exponent $5/12$ is itself conditional on extrapolating the adopted
$\Omegaa\propto\fa^{7/6}$ abundance law. In the asymptotic constant-mass
regime, $\Omegaa\propto\fa^{3/2}$ at fixed $\thetai$, so the abundance
condition instead gives $\thetai\propto\fa^{-3/4}$ and hence
$\HI^{\rm max}\propto\fa^{1/4}$~\cite{Visinelli:2009zm,Marsh:2015xka}.
The dependence on $\fa$ is therefore even weaker in that regime.

This weak decay constant dependence is the essential point. Within the
adopted fit, increasing $\fa$ by more than six orders of magnitude, from
$10^{12}\GeV$ to the reduced Planck scale, raises the allowed $\HI$ by fewer
than three orders of magnitude. Thus, even a near Planckian decay constant
does not reconcile the minimal pre-inflationary QCD axion scenario with
conventional high-scale inflation~\cite{Hertzberg:2008wr,Beltran:2006sq}.
This is particularly relevant because decay constants
$\fa\sim10^{15}$--$10^{16}\GeV$ commonly arise in string axion
constructions~\cite{Svrcek:2006yi,Fox:2004kb}. Evading the conclusion
therefore requires relaxing at least one assumption entering the minimal
bound, as discussed in Sec.~\ref{sec:loopholes}.

\section{Primordial tensor modes and the Lyth bound}
\label{sec:lyth}

\subsection{The tensor scale and the axion isocurvature bound on
\texorpdfstring{$r$}{r}}
\label{subsec:axiontensor}

Let $\mathcal P_t$ denote the dimensionless primordial tensor power spectrum
and let $\PRR$ denote the curvature power spectrum, normalized as
$\PRR(k_*)=\As$ at the pivot scale. In Einstein gravity, assuming the standard
inflationary tensor vacuum, the tensor power is, to leading order in slow
roll, determined by the Hubble scale at horizon exit,
\begin{equation}
  \mathcal P_t(k_*)
  =
  \frac{2\HI^2}{\pi^2\Mpl^2},
  \qquad
  r(k_*)
  \equiv
  \frac{\mathcal P_t(k_*)}{\PRR(k_*)}
  =
  \frac{\mathcal P_t(k_*)}{\As}.
  \label{eq:tensorH}
\end{equation}
This relation does not require the inflaton itself to generate the observed
scalar curvature perturbation. It does, however, assume the standard
normalization and propagation of primordial tensor modes.

For a light axion during inflation, the same Hubble scale controls the axion
fluctuation. Eliminating $\HI$ between Eqs.~\eqref{eq:PIIexact} and
\eqref{eq:tensorH}, and imposing a direct observational upper limit on the CDI
amplitude, gives
\begin{equation}
  r
  <
  8\,
  \frac{\PII^{\rm max}(k_*)}{\As}
  \left[
    \frac{\fI}
         {\gamma_a\,|\Ttheta\Gresp(\thetai)|\,\Mpl}
  \right]^2.
  \label{eq:axiontensorPII}
\end{equation}
For a fixed parameter point, the same result may be rewritten in terms of
$\betaiso$ using
$\PII(k_*)=\As\betaiso(k_*)/[1-\betaiso(k_*)]$. As in
Sec.~\ref{subsec:power}, however, we use the directly reported limits on
$\PII$ for numerical estimates.

For minimal transfer, $\Ttheta=1$, the harmonic result
$|\Gresp|=2/|\thetai|$ reduces the bound to
\begin{equation}
  r
  <
  2\,
  \frac{\PII^{\rm max}(k_*)}{\As}
  \left(
    \frac{\fI|\thetai|}
         {\gamma_a\Mpl}
  \right)^2.
  \label{eq:axiontensorharmonicPII}
\end{equation}

Equation~\eqref{eq:axiontensorPII} expresses the fact that primordial tensors
and axion isocurvature are sourced by the same inflationary Hubble scale and
therefore cannot be varied independently once the axion sector is
specified~\cite{Hertzberg:2008wr,Marsh:2014qoa}. For fixed axion parameters, the
isocurvature limit places an upper bound on $r$. Conversely, a future tensor
detection, combined with information about the axion abundance and the
inflationary normalization $\fI$, would test the corresponding
pre-inflationary axion scenario.

For example, a tensor signal at $r=10^{-3}$ would correspond to
\begin{equation}
  \HI
  =
  \pi\Mpl
  \left(
    \frac{\As r}{2}
  \right)^{1/2}
  \simeq
  7.9\times10^{12} \, \GeV \,,
  \label{eq:HrExample}
\end{equation}
where we used $\As=2.10\times10^{-9}$. This value is more than two orders of
magnitude above the most permissive ceiling found in
Sec.~\ref{subsec:Hmax}. Thus, under the tensor assumptions stated above and
within the benchmark abundance fit, a detection near $r\sim10^{-3}$ would
exclude the minimal pre-inflationary QCD axion scenario in which
$\fI=\fa$, $\Ttheta=1$, $\Omegaa=\OmegaDM$, and $\fa\leq\Mpl$. Such a
detection would not exclude more general axion constructions with
$\fI>\fa$, a non-negligible axion mass during inflation, a subdominant axion
fraction, or a nontrivial superhorizon transfer history.

It is useful to record the general conversion between $\HI$ and the tensor
observables. Using $\As=2.10\times10^{-9}$ and
$\Mpl=2.435\times10^{18}\GeV$,
\begin{align}
  r
  &=
  \frac{2\HI^2}
       {\pi^2\Mpl^2\As}
  =
  1.63\times10^{-15}
  \left(
    \frac{\HI}{10^7\GeV}
  \right)^2,
  \label{eq:rH}
  \\
  V_*^{1/4}
  &\simeq
  \left(
    3\Mpl^2\HI^2
  \right)^{1/4}
  =
  6.49\times10^{12}\GeV
  \left(
    \frac{\HI}{10^7\GeV}
  \right)^{1/2}.
  \label{eq:Vscale}
\end{align}
The second relation assumes that the inflationary energy density is dominated
by the potential, as in slow-roll inflation.

If the observed curvature perturbation is generated by a canonical
single-field slow-roll inflaton, one additionally has, to leading order in
slow roll,
\begin{equation}
  \epsilon_H
  \simeq
  \epsilon_V
  \simeq
  \frac{\HI^2}
       {8\pi^2\Mpl^2\As}
  =
  1.02\times10^{-16}
  \left(
    \frac{\HI}{10^7\GeV}
  \right)^2,
  \label{eq:epsilonH}
\end{equation}
together with $r\simeq16\epsilon_H$. These relations do not apply unchanged
when a spectator field generates a significant part of the observed curvature
perturbation, even though Eq.~\eqref{eq:tensorH} remains valid.

For the representative range $\HI\sim10^7\text{--}10^8\GeV$, one obtains
$r\sim10^{-15}\text{--}10^{-13}$ and
$V_*^{1/4}\sim10^{13}\GeV$. The tensor amplitude is then many orders of
magnitude below the current direct limit $r_{0.05}<0.036$ at $95\%$
C.L.~\cite{BICEP:2021xfz,BICEPKeck:2024stm}, and below the forecast
sensitivities of next-generation CMB polarization
experiments~\cite{LiteBIRD:2022cnt}. Thus, the portion of the minimal
axion parameter space with an order-one initial angle predicts an effectively
unobservable primordial tensor signal.

\subsection{The Lyth bound at low inflationary scales}
\label{subsec:lythlow}

Before specializing further to the axion scenario, it is useful to recall the
relation between the tensor amplitude and the inflaton field range. For a
canonical single-field slow-roll model in which the inflaton generates the
observed curvature perturbation,
\begin{equation}
  \frac{1}{\Mpl}
  \left|
    \frac{\dd\phi}{\dd N}
  \right|
  =
  \sqrt{2\epsilon_H}
  \simeq
  \sqrt{\frac{r}{8}}.
  \label{eq:lythdiff}
\end{equation}
For monotonic field evolution over an interval
$N\in[N_1,N_2]$, with $\Delta N\equiv N_2-N_1>0$, the corresponding
excursion is
\begin{equation}
  \frac{|\Delta\phi|}{\Mpl}
  \simeq
  \int_{N_1}^{N_2}
  \sqrt{\frac{r(N)}{8}}\,
  \dd N.
  \label{eq:lythintegral}
\end{equation}
This is the basis of the Lyth bound: an appreciable tensor amplitude sustained
over a sufficiently long interval requires a correspondingly large field
excursion~\cite{Lyth:1996im}. Stronger statements relating
$r\gtrsim10^{-3}$ to a super Planckian total excursion require additional
assumptions about the evolution of the slow-roll parameters and the shape of
the inflationary potential~\cite{Efstathiou:2005tq}. Generalizations of the
field-range bound beyond canonical slow roll can be formulated using the
effective theory of single-field inflation, and a weaker bound persists under
the null energy condition~\cite{Baumann:2011ws}.

At low inflationary scales, the associated tensor amplitude is tiny and the
usual Lyth lower bound is correspondingly weak. For slowly varying $\HI$, the
horizon crossing relation $k=aH$ gives
$\Delta N_{\rm CMB}\simeq\ln(k_{\max}/k_{\min})$. Taking
$k_{\min}\sim10^{-4}\,\mathrm{Mpc}^{-1}$ and
$k_{\max}\sim0.2\,\mathrm{Mpc}^{-1}$ gives
$\Delta N_{\rm CMB}\simeq7.6$, for which we use the representative
normalization $\Delta N_{\rm CMB}=8$. Approximating $r(N)$ by its pivot-scale
value across this window, Eqs.~\eqref{eq:rH} and
\eqref{eq:lythintegral} give
\begin{equation}
  \frac{|\Delta\phi_{\rm CMB}|}{\Mpl}
  \simeq
  1.14\times10^{-7}
  \left(
    \frac{\Delta N_{\rm CMB}}{8}
  \right)
  \left(
    \frac{\HI}{10^7\GeV}
  \right).
  \label{eq:fieldrangeNumeric}
\end{equation}
Under this approximation, the axion isocurvature limit on $\HI$ becomes an
upper limit on the local inflaton excursion. For the representative
order-one angle benchmark, $\HI<2.2\times10^7\GeV$ implies
$|\Delta\phi_{\rm CMB}|/\Mpl\lesssim2.5\times10^{-7}$ for
$\Delta N_{\rm CMB}=8$. Even at the prior-dependent, most permissive benchmark
ceiling, $\HI<1.25\times10^{10}\GeV$, the corresponding result is
$|\Delta\phi_{\rm CMB}|/\Mpl\lesssim1.43\times10^{-4}$, provided the tensor
amplitude remains approximately constant across this interval.

Several consequences follow. First, within the canonical single-field setup,
the observable scalar spectrum is generated over a very small interval in
field space. The tilt, its running, and any localized features constrain only
the local structure of the inflationary potential over
$|\Delta\phi_{\rm CMB}|\ll\Mpl$. CMB observations alone therefore cannot
reconstruct the potential far beyond this neighborhood.

Second, the axion-compatible low-scale region naturally favors very small
field motion over the observable window. Models based on a locally flat
plateau, a hilltop, or an approximate inflection point can realize such
behavior, whereas standard monomial large-field models generally predict a
much larger tensor amplitude. This does not imply that every viable low-scale
model has a sub-Planckian total field excursion, because the inflaton can
evolve more rapidly outside the observable window.

Third, the field-range statement is intrinsically local. It constrains the
motion while the relevant modes leave the horizon, not the total trajectory
from the beginning to the end of inflation. The tensor amplitude may evolve
outside the observable window, particularly near the end of inflation.
Moreover, the number of $e$-folds between pivot exit and the end of inflation
depends on the reheating history and can differ from its conventional
high-scale estimate. In more general single-field theories, both the
field-range statement and the relation between tensor and scalar fluctuations
must be reformulated~\cite{Baumann:2011ws}. In multifield theories, the
relevant local quantity is the trajectory length computed with the
field-space metric, and entropy-to-curvature transfer further modifies its
connection to $r$. Equation~\eqref{eq:fieldrangeNumeric} should therefore be
interpreted as a local canonical single-field estimate rather than as a
model-independent bound on the complete inflationary trajectory.

\subsection{The Lyth bound with an axion isocurvature constraint}
\label{subsec:axionlyth}

Within canonical single-field slow-roll inflation, the local Lyth relation
can be combined with the axion isocurvature bound on the tensor amplitude.
Using Eq.~\eqref{eq:axiontensorPII}, one finds at the pivot scale
\begin{equation}
  \left.
  \frac{1}{\Mpl}
  \left|
    \frac{\dd\phi}{\dd N}
  \right|
  \right|_{*}
  <
  \sqrt{
    \frac{\PII^{\rm max}(k_*)}{\As}
  }\,
  \frac{\fI}
       {\gamma_a\,|\Ttheta\Gresp(\thetai)|\,\Mpl}.
  \label{eq:axionlythlocalPII}
\end{equation}
Equation~\eqref{eq:axionlythlocalPII} bounds the local inflaton speed at the
pivot. Unlike the conventional Lyth bound, it does not directly constrain the
total field excursion.

For the nearly scale-invariant CDI spectrum considered here, if the
background quantities entering Eq.~\eqref{eq:axionlythlocalPII} vary slowly
enough that its right hand side is approximately constant across the
observable CMB window, the local bound may be integrated to give
\begin{equation}
  \frac{|\Delta\phi_{\rm CMB}|}{\Mpl}
  \lesssim
  \Delta N_{\rm CMB}
  \sqrt{
    \frac{\PII^{\rm max}(k_*)}{\As}
  }\,
  \frac{\fI}
       {\gamma_a\,|\Ttheta\Gresp(\thetai)|\,\Mpl}.
  \label{eq:axionlythPII}
\end{equation}

For harmonic misalignment with minimal transfer, $\Ttheta=1$ and
$|\Gresp|=2/|\thetai|$, so Eq.~\eqref{eq:axionlythPII} becomes
\begin{equation}
  \frac{|\Delta\phi_{\rm CMB}|}{\Mpl}
  \lesssim
  \frac{\Delta N_{\rm CMB}}{2}
  \sqrt{
    \frac{\PII^{\rm max}(k_*)}{\As}
  }\,
  \frac{\fI|\thetai|}
       {\gamma_a\Mpl}.
  \label{eq:axionlythharmPII}
\end{equation}

This result complements rather than replaces the conventional Lyth bound.
The usual Lyth argument relates a measured tensor amplitude to a lower bound
on the field excursion when the tensor signal is sustained over an interval.
By contrast, axion isocurvature provides an upper limit on the tensor
amplitude and therefore on the local inflaton speed. Only when the relevant
quantities vary slowly over the observable window does this local constraint
translate into the approximate field range bound in
Eq.~\eqref{eq:axionlythPII}.

In the minimal benchmark, Eq.~\eqref{eq:fieldrangeNumeric} gives
$|\Delta\phi_{\rm CMB}|/\Mpl\lesssim2.5\times10^{-7}$ for the representative
order-one angle case and $|\Delta\phi_{\rm CMB}|/\Mpl\lesssim1.43\times10^{-4}$ at the most permissive
benchmark ceiling, taking $\Delta N_{\rm CMB}=8$. Thus, the minimal
pre-inflationary QCD axion scenario predicts both an unobservably small
primordial tensor signal and very little inflaton motion while observable CMB
scales leave the horizon, provided the curvature perturbation is generated by
a canonical slow-roll inflaton and the slow-variation assumptions above hold.
This conclusion applies only to the observable window: the field may evolve
much more rapidly outside this interval, so the result should not be
interpreted as a global upper bound on the total inflationary trajectory.

\subsection{The hierarchy of potential derivatives}
\label{subsec:slopecurvature}
The small tensor amplitude constrains the local slope of the inflaton
potential, while the measured scalar tilt constrains its curvature. Together,
they imply a pronounced hierarchy between the first two dimensionless
derivatives of the potential. This discussion assumes canonical single-field
slow-roll inflation in which the inflaton generates the observed curvature
perturbation. At leading order,
\begin{equation}
  \epsilon_V
  =
  \frac{\Mpl^2}{2}
  \left(
    \frac{V'}{V}
  \right)^2,
  \qquad
  \eta_V
  =
  \Mpl^2
  \frac{V''}{V},
  \qquad
  n_s-1
  =
  -6\epsilon_V+2\eta_V.
  \label{eq:slowroll}
\end{equation}
It is convenient to define
\begin{equation}
  \lambda_1
  \equiv
  \Mpl\frac{V'}{V},
  \qquad
  \lambda_2
  \equiv
  \Mpl^2\frac{V''}{V},
  \label{eq:lambdadefs}
\end{equation}
so that $\epsilon_V=\lambda_1^2/2$ and $\eta_V=\lambda_2$.

Using the observed scalar amplitude, the dimensionless potential slope is
\begin{equation}
  |\lambda_1|
  =
  \sqrt{2\epsilon_V}
  \simeq
  \frac{\HI}
       {2\pi\Mpl\sqrt{\As}}
  =
  1.43\times10^{-8}
  \left(
    \frac{\HI}{10^7\GeV}
  \right).
  \label{eq:lambda1}
\end{equation}
Throughout the axion-compatible region,
$\epsilon_V\ll|\eta_V|$, so the scalar tilt gives
\begin{equation}
  \lambda_2
  =
  \eta_V
  \simeq
  \frac{n_s-1}{2}.
  \label{eq:lambda2}
\end{equation}
For example, Planck 2018 gives
$n_s=0.9649\pm0.0042$, corresponding at leading order to
\begin{equation}
  \eta_V
  \simeq
  -0.0176\pm0.0021
  \qquad
  \text{(Planck 2018)},
  \label{eq:etaPlanck}
\end{equation}
while the representative ACT DR6 P-ACT-LB combination gives
\begin{equation}
  \eta_V
  \simeq
  -0.0129\pm0.0017
  \qquad
  \text{(ACT DR6)},
  \label{eq:etaACT}
\end{equation}
from $n_s=0.9743\pm0.0034$~\cite{Planck:2018jri,AtacamaCosmologyTelescope:2025blo}. These results use different dataset
combinations and are quoted only to illustrate the range of curvature
preferred by current analyses.

\begin{figure}[t]
\centering
\IfFileExists{figures/slope_curvature_hierarchy.pdf}{%
  \includegraphics[width=0.91\textwidth]
  {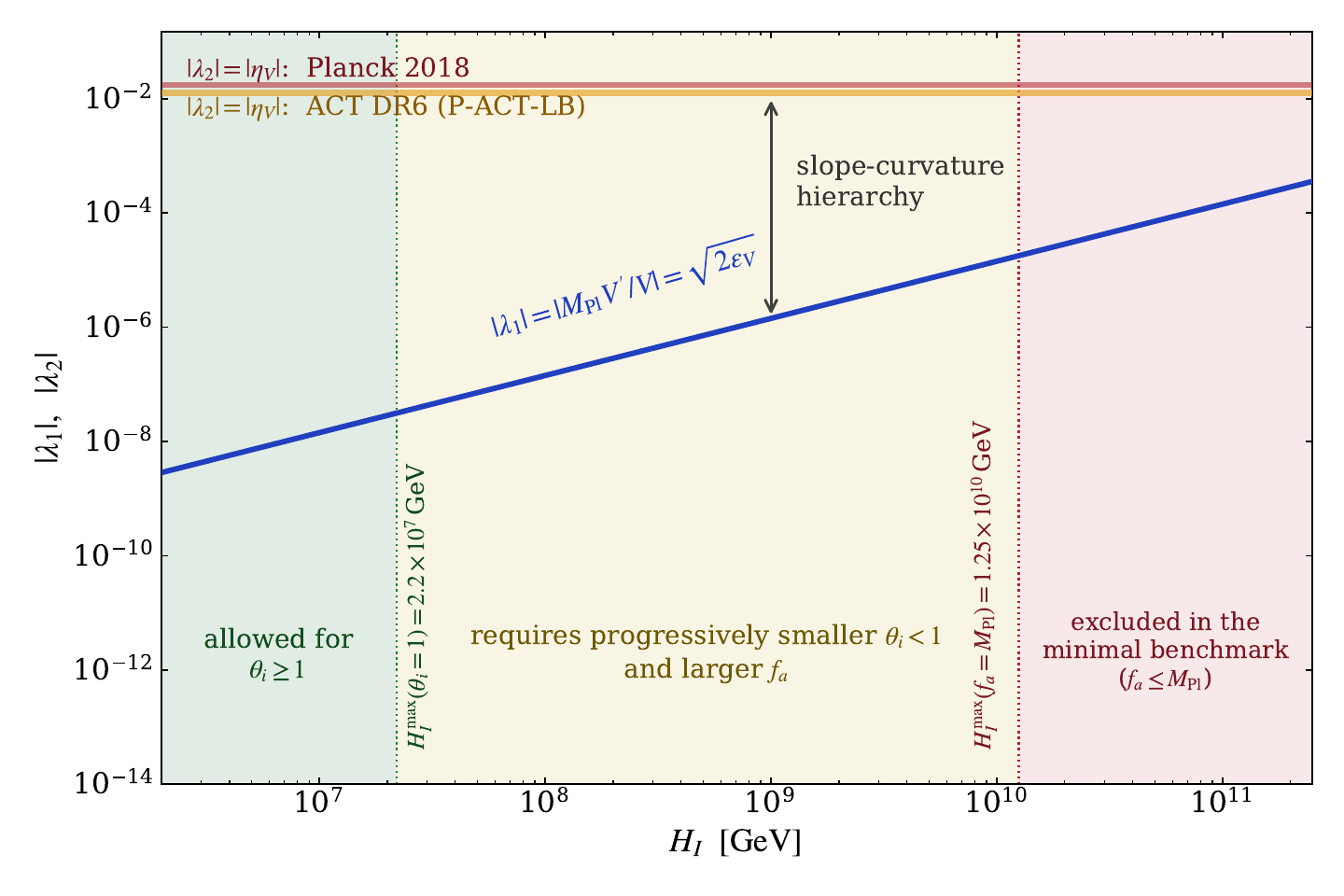}
}{%
  \fbox{%
    \parbox{0.85\textwidth}{%
      \centering
      Missing file
      \texttt{slope\_curvature\_hierarchy.pdf}.
    }%
  }
}
\caption{Hierarchy between the first two dimensionless potential derivatives
in canonical single-field slow-roll inflation. The observed scalar amplitude
fixes $|\lambda_1|=|\Mpl V'/V|$ as a function of $\HI$, while the scalar tilt
fixes $|\lambda_2|=|\Mpl^2V''/V|$ near $10^{-2}$. The horizontal bands show
representative Planck 2018 and ACT DR6 determinations. The shaded interval
spans the axion-motivated range between the order-one angle benchmark,
$\HI\simeq2.2\times10^7\GeV$, and the prior-dependent $\fa\leq\Mpl$ ceiling,
$\HI\simeq1.25\times10^{10}\GeV$.}
\label{fig:derivativehierarchy}
\end{figure}

Across the benchmark interval
$\HI\simeq10^7$--$10^{10}\GeV$, the resulting hierarchy is
\begin{equation}
  \frac{\epsilon_V}{|\eta_V|}
  \sim
  10^{-14}\text{--}10^{-8}.
  \label{eq:epsEtaHierarchy}
\end{equation}
The inflaton potential must therefore have an extremely small dimensionless
slope while retaining a negative dimensionless curvature of order $10^{-2}$.
This hierarchy is shown in Fig.~\ref{fig:derivativehierarchy}.

Maintaining this hierarchy is nontrivial in a generic effective theory.
In the absence of a protective symmetry or other special structure,
Planck-suppressed couplings between the inflaton and the inflationary vacuum
energy can generate a mass correction of order
\begin{equation}
  \delta m_\phi^2
  \sim
  \frac{V}{\Mpl^2}
  \sim
  \HI^2.
  \label{eq:etamasscorrection}
\end{equation}
Such a correction produces
\begin{equation}
  \delta\eta_V
  \sim
  \frac{\delta m_\phi^2\Mpl^2}{V}
  \sim
  \frac{\delta m_\phi^2}{3\HI^2}
  =
  \mathcal O(1),
  \label{eq:etacorrection}
\end{equation}
in conflict with the slow-roll requirement
$|\eta_V|\ll1$. This is the usual eta problem, particularly familiar in
supergravity realizations of inflation~\cite{Copeland:1994vg,Dine:1995uk}.

Lowering the overall normalization of a potential with fixed shape does not
resolve this problem. Under
$V(\phi)\rightarrow cV(\phi)$, the quantities $V$, $V'$, and $V''$ scale
together, leaving $V'/V$, $V''/V$, and hence the potential slow-roll
parameters unchanged. A viable low-scale model must therefore possess an
appropriately flat local shape, often protected by a symmetry or produced by
a controlled structure in the potential. Simply rescaling a successful
high-scale potential is insufficient.

The same derivative hierarchy affects the interpretation of the running.
Define
\begin{equation}
  \xi_V^2
  \equiv
  \Mpl^4
  \frac{V'V'''}{V^2},
  \qquad
  \lambda_3
  \equiv
  \Mpl^3
  \frac{V'''}{V}.
  \label{eq:xidef}
\end{equation}
At leading order~\cite{Kosowsky:1995aa},
\begin{equation}
  \alpha_s
  \equiv
  \frac{\dd n_s}{\dd\ln k}
  =
  16\epsilon_V\eta_V
  -
  24\epsilon_V^2
  -
  2\xi_V^2,
  \qquad
  \xi_V^2
  =
  \lambda_1\lambda_3.
  \label{eq:running}
\end{equation}
Since $\epsilon_V$ is extremely small throughout the relevant region,
\begin{equation}
  \xi_V^2
  \simeq
  -\frac{\alpha_s}{2},
  \qquad
  \lambda_3
  \simeq
  -\frac{\alpha_s}{2\lambda_1}.
  \label{eq:runningLowScale}
\end{equation}
Because $|\lambda_1|$ is tiny, even a modest nonzero running corresponds to a
large dimensionless third derivative. For illustration,
$|\alpha_s|\sim10^{-3}$ would imply
\begin{equation}
  |\xi_V^2|
  \sim
  5\times10^{-4},
  \qquad
  |\lambda_3|
  \sim
  3.5\times10^4
  \left(
    \frac{10^7\GeV}{\HI}
  \right).
  \label{eq:lambda3estimate}
\end{equation}
Current measurements remain consistent with negligible running. For example,
the ACT DR6 extended analysis finds $\alpha_s = 0.0062\pm0.0052$~\cite{AtacamaCosmologyTelescope:2025nti}.

Low-scale inflation therefore makes the running especially sensitive to
localized structure and higher derivatives of the inflaton potential. A large
value of $|\lambda_3|$ is not by itself inconsistent with slow roll, because
the quantity entering the slow-roll hierarchy is the product
$\xi_V^2=\lambda_1\lambda_3$. It does, however, show that a Taylor expansion
organized solely by the magnitudes of the individual dimensionless
derivatives $\lambda_n$ need not be well behaved in the low-scale regime. A
viable model must instead realize the required derivative hierarchy through a
symmetry, an approximate inflection point, a plateau, or another controlled
functional structure.

\section{Reheating, the \texorpdfstring{$e$}{e}-fold count, and PQ
non-restoration}
\label{sec:reheating}
The preceding section translated the inflationary Hubble scale and the
measured scalar tilt into local constraints on the inflaton potential.
Embedding these constraints in a complete inflationary model requires
specifying the post-inflationary expansion history. Reheating enters in two
distinct ways: it determines the number of $e$-folds $N_*$ between
pivot-scale horizon exit and the end of inflation, and it helps determine
whether the Peccei--Quinn (PQ) symmetry remains broken after inflation. For
commonly studied plateau and hilltop potentials, lowering the inflationary
scale or prolonging a reheating phase with $w_{\rm reh}<1/3$ reduces $N_*$ and
therefore generally lowers the predicted scalar tilt.

\subsection{The \texorpdfstring{$e$}{e}-fold count at low inflationary scales}
\label{subsec:Nstar}

The relation between the CMB pivot scale and the inflaton potential depends on
the expansion history between the end of inflation and the onset of radiation
domination. Approximating the reheating epoch by a constant effective equation
of state $w_{\rm reh}$ and assuming the standard entropy-conserving thermal
history thereafter, one obtains~\cite{Liddle:2003as,Planck:2018jri}
\begin{align}
  N_*
  \simeq {}&
  66.9
  -
  \ln\!\left(
    \frac{k_*}{a_0H_0}
  \right)
  +
  \frac{1}{4}
  \ln\!\left(
    \frac{V_*^2}
         {\Mpl^4\rho_{\rm end}}
  \right)
  +
  \frac{1-3w_{\rm reh}}
       {12(1+w_{\rm reh})}
  \ln\!\left(
    \frac{\rho_{\rm RH}}
         {\rho_{\rm end}}
  \right)
  -
  \frac{1}{12}\ln g_{\rm RH}.
  \label{eq:Nstar}
\end{align}
Here $V_*$ is the potential energy at pivot-scale horizon exit,
$\rho_{\rm end}$ is the total energy density at the end of inflation, and
$\rho_{\rm RH}$ is the radiation energy density when reheating is completed.
We have identified the effective energy and entropy degrees of freedom at
that time and denoted their common value by $g_{\rm RH}$. The reheating
temperature is defined through
\begin{equation}
  \rho_{\rm RH}
  =
  \frac{\pi^2}{30}\,
  g_{\rm RH}T_{\rm RH}^4.
  \label{eq:rhoRH}
\end{equation}

For the pivot scale $k_*=0.05\,{\rm Mpc}^{-1}$, standard present-day
cosmological parameters, $g_{\rm RH}=106.75$, nearly constant energy density
$V_*\simeq\rho_{\rm end}\simeq3\Mpl^2\HI^2$, and instantaneous reheating,
$\rho_{\rm RH}=\rho_{\rm end}$, Eq.~\eqref{eq:Nstar} gives
\begin{equation}
  N_*^{\rm inst}
  \simeq
  48.3
  +
  \frac{1}{2}
  \ln\!\left(
    \frac{\HI}{10^7\GeV}
  \right).
  \label{eq:Ninstant}
\end{equation}
The numerical constant retains an uncertainty of order one $e$-fold from the
dynamics near the end of inflation, deviations of $\rho_{\rm end}$ from
$V_*$, changes in the relativistic degrees of freedom, and the detailed
transition to radiation domination.

For the representative interval
$\HI\sim10^7$--$10^{10}\GeV$, instantaneous reheating therefore gives
approximately $N_*^{\rm inst}\simeq48$--$52$, with still lower inflationary
scales giving smaller values. This is below the frequently adopted estimate
$N_*\simeq55$--$60$, which is more typical of higher inflationary scales and
efficient reheating.

For a matter-like reheating phase, $w_{\rm reh}=0$, define the instantaneous
reheating temperature by
\begin{equation}
  \rho_{\rm end}
  =
  \frac{\pi^2}{30}\,
  g_{\rm RH}
  \left(T_{\rm RH}^{\rm inst}\right)^4.
  \label{eq:TRHinst}
\end{equation}
The reheating contribution to the $e$-fold count can then be written as
\begin{equation}
  \Delta N_*
  \simeq
  \frac{1}{3}
  \ln\!\left(
    \frac{T_{\rm RH}}
         {T_{\rm RH}^{\rm inst}}
  \right),
  \label{eq:DeltaNTRH}
\end{equation}
up to the weak dependence on the relativistic degrees of freedom. Thus, each
decade by which $T_{\rm RH}$ lies below its instantaneous value reduces $N_*$
by
\begin{equation}
  |\Delta N_*|
  \simeq
  \frac{1}{3}\ln 10
  \simeq
  0.77.
\end{equation}
More generally, reheating with $w_{\rm reh}<1/3$ lowers $N_*$ relative to
instantaneous reheating. A phase with $w_{\rm reh}=1/3$ leaves the matching
independent of $\rho_{\rm RH}$ at this order, whereas $w_{\rm reh}>1/3$ can
increase $N_*$ as $T_{\rm RH}$ is lowered. The latter possibility requires a
sufficiently stiff post-inflationary equation of state and is therefore
model dependent.

\subsection{PQ non-restoration and the maximum temperature}
\label{subsec:PQrestore}
The pre-inflationary PQ-breaking scenario requires the symmetry to remain
broken throughout the post-inflationary evolution. The commonly imposed
condition $T_{\rm RH}<\fa$ is not by itself sufficient: during a prolonged
reheating phase, the radiation bath can reach a maximum temperature
$T_{\max}$ substantially higher than the temperature $T_{\rm RH}$ at which
radiation domination begins. Moreover, the relevant restoration temperature
is determined by the finite temperature dynamics of the PQ sector and need
not coincide with $\fa$.

In the instantaneous reheating limit, the inflationary energy density is
converted directly into radiation, and the maximum and reheating temperatures
coincide:
\begin{equation}
  T_{\rm inst}
  =
  \left(
    \frac{30\rho_{\rm end}}
         {\pi^2g_{\rm RH}}
  \right)^{1/4}
  \simeq
  2.7\times10^{12}\GeV
  \left(
    \frac{H_{\rm end}}{10^7\GeV}
  \right)^{1/2}
  \left(
    \frac{g_{\rm RH}}{106.75}
  \right)^{-1/4},
  \label{eq:Tinst}
\end{equation}
where $\rho_{\rm end}\simeq3\Mpl^2H_{\rm end}^2$ has been used. For the
order of magnitude estimates below, we take $H_{\rm end}\sim\HI$. The precise
relation is model dependent.

If reheating instead proceeds perturbatively through an extended
inflaton-dominated phase, radiation is produced before radiation domination.
Under the conventional assumption that the inflaton decay products thermalize
sufficiently rapidly, the maximum temperature for a matter-like oscillating
condensate scales as~\cite{Giudice:2000ex, Garcia:2017tuj, Garcia:2020eof, Garcia:2020wiy}
\begin{equation}
  T_{\max}
  \simeq
  C_{\max}
  \left(
    T_{\rm RH}^2H_{\rm end}\Mpl
  \right)^{1/4},
  \qquad
  C_{\max}=\mathcal O(0.1\text{--}1).
  \label{eq:Tmax}
\end{equation}
Here $C_{\max}$ depends on the reheating convention, the inflaton decay history, and the effective relativistic degrees of freedom. Equation~\eqref{eq:Tmax} should be regarded as a rapid thermalization benchmark: delayed thermalization can substantially lower the effective maximum temperature relevant for finite-density corrections to the PQ potential~\cite{Mukaida:2015ria}. Within the rapid thermalization approximation,
\begin{equation}
  \frac{T_{\max}}{T_{\rm RH}}
  \simeq
  C_{\max}
  \left(
    \frac{H_{\rm end}\Mpl}
         {T_{\rm RH}^2}
  \right)^{1/4},
  \label{eq:TmaxOverTRH}
\end{equation}
so lowering $T_{\rm RH}$ reduces $T_{\max}$ only as $T_{\rm RH}^{1/2}$ at fixed $H_{\rm end}$.

When the PQ sector is in thermal contact with the radiation bath, a
sufficient condition for avoiding thermal restoration is
\begin{equation}
  T_{\max}\lesssim T_{\rm PQ},
  \label{eq:PQthermalcondition}
\end{equation}
where $T_{\rm PQ}$ is the critical temperature of the PQ sector. Its value is
set by the radial potential and the finite-density corrections induced by
particles coupled to the PQ field~\cite{Mukaida:2015ria}. Schematically, $T_{\rm PQ}\sim \sqrt{\lambda_\Phi/c_T}\, v_{\rm PQ}$, where $\lambda_\Phi$ characterizes the zero-temperature radial potential,
$c_T$ is the thermal-mass coefficient, and $v_{\rm PQ}$ is the radial vacuum
expectation value. Consequently, $T_{\rm PQ}=\mathcal O(\fa)$ only when the
relevant coupling ratio and the relation between $v_{\rm PQ}$ and $\fa$ are
both of order unity. Weak coupling to the bath can increase the restoration
temperature or prevent the PQ sector from thermalizing altogether, in which
case a constraint based solely on the visible-sector temperature does not
apply.

Within the rapid thermalization benchmark, combining
Eqs.~\eqref{eq:Tmax} and \eqref{eq:PQthermalcondition} gives
\begin{equation}
  T_{\rm RH}
  \lesssim
  \frac{T_{\rm PQ}^2}
       {C_{\max}^2\sqrt{H_{\rm end}\Mpl}}.
  \label{eq:TRHPQbound}
\end{equation}
For the natural angle dark matter benchmark,
$\fa\simeq9\times10^{11}\GeV$ and
$\HI\simeq2.2\times10^7\GeV$, the instantaneous-reheating limit gives
\begin{equation}
  T_{\rm inst}
  \simeq
  4.0\times10^{12}\GeV
  \left(
    \frac{g_{\rm RH}}{106.75}
  \right)^{-1/4},
  \label{eq:TinstNaturalAngle}
\end{equation}
where $H_{\rm end}\sim\HI$ has been assumed. This temperature exceeds $\fa$
by a factor of a few. Therefore, if $T_{\rm PQ}\sim\fa$ and the PQ sector is
efficiently coupled to the thermal bath, instantaneous reheating would be
expected to restore the symmetry.

Taking $C_{\max}=1$, $T_{\rm PQ}=9\times10^{11}\GeV$, and
$H_{\rm end}=2.2\times10^7\GeV$, Eq.~\eqref{eq:TRHPQbound} gives
\begin{equation}
  T_{\rm RH}
  \lesssim
  1.1\times10^{11}\GeV
  \left(
    \frac{T_{\rm PQ}}{9\times10^{11}\GeV}
  \right)^2
  \left(
    \frac{C_{\max}}{1}
  \right)^{-2}
  \left(
    \frac{H_{\rm end}}{2.2\times10^7\GeV}
  \right)^{-1/2}.
  \label{eq:TRHPQNumeric}
\end{equation}
Saturating this limit during matter-like reheating gives
$\Delta N_*\simeq-1.2$ relative to instantaneous reheating, while a lower
$T_{\rm RH}$ produces a larger reduction. Under these benchmark assumptions,
thermal PQ non-restoration therefore imposes an additional constraint and
shifts $N_*$ in the same downward direction as the low inflationary scale.
The quantitative restriction remains model dependent through
$T_{\rm PQ}$, $C_{\max}$, the thermalization history, and the coupling of the
PQ sector to the radiation bath.

The thermal condition alone does not guarantee that the PQ symmetry remains
broken. Coherent oscillations of the PQ radial field can undergo parametric
amplification and restore the symmetry nonthermally even when
$T_{\max}<T_{\rm PQ}$~\cite{Kofman:1995fi,Kawasaki:2013ae,Harigaya:2015hha}. Avoiding restoration
therefore requires control of both the thermal bath and the post-inflationary
dynamics of the PQ field.

Finally, reheating cannot be delayed arbitrarily. In conventional reheating
scenarios, successful Big Bang nucleosynthesis requires approximately
$T_{\rm RH}\gtrsim4\,{\rm MeV}$ at 95\% C.L.~\cite{Hannestad:2004px,deSalas:2015glj}. In the scenarios considered here,
however, the reheating-induced reduction of $N_*$ and its effect on the
predicted scalar tilt generally become relevant long before this absolute
lower limit is approached.

\subsection{A three-way consistency problem for axions and low-scale inflation}
\label{subsec:threeway}
The reheating-dependent $e$-fold count feeds directly into the predicted
scalar tilt. Plateau models with the usual large-$N_*$ attractor behavior
give
\begin{equation}
  n_s
  \simeq
  1-\frac{2}{N_*}
  \label{eq:plateauns}
\end{equation}
at leading order~\cite{Starobinsky:1980te,Kallosh:2013yoa}. By contrast,
small-field hilltop models of the form
\begin{equation}
  V(\phi)
  =
  V_0
  \left[
    1-
    \left(
      \frac{\phi}{\mu}
    \right)^p
    +\cdots
  \right],
  \qquad
  p>2,
  \label{eq:hilltoppotential}
\end{equation}
give, in the usual small-field and large-$N_*$ regime,
\begin{equation}
  n_s
  \simeq
  1-
  \frac{2(p-1)}
       {(p-2)N_*}
  \label{eq:hilltopns}
\end{equation}
at leading order~\cite{Boubekeur:2005zm}. Since
\begin{equation}
  \frac{2(p-1)}{p-2}>2
  \qquad
  \text{for finite }p>2,
\end{equation}
the hilltop prediction is redder than the plateau result at fixed $N_*$,
approaching it only in the formal limit $p\to\infty$.

For the low-scale range $N_*\simeq48$--$52$, the plateau prediction is $n_s^{\rm plateau} \simeq 0.9583\text{--}0.9615$.
This range is compatible with, but lies below the central value of, the Planck 2018 determination
$n_s=0.9649\pm0.0042$. It is approximately $3.8$--$4.7\,\sigma$ below the representative ACT DR6 P-ACT-LB result
$n_s=0.9743\pm0.0034$, when only the quoted observational uncertainty is
used~\cite{Planck:2018jri,AtacamaCosmologyTelescope:2025blo}. Finite-$p$
hilltop models predict still smaller values of $n_s$ and therefore face at
least as much tension. For example, $p=4$ gives
\begin{equation}
  n_s
  \simeq
  1-\frac{3}{N_*}
  \simeq
  0.9375\text{--}0.9423
\end{equation}
over the same interval.

When thermal PQ non-restoration is enforced by delaying reheating, the
resulting shift in $N_*$ exacerbates this tension for
$w_{\rm reh}<1/3$. Lowering $T_{\rm RH}$ then decreases $N_*$ and shifts both
the plateau and hilltop predictions toward still redder spectra. Thus, in
models where preventing PQ restoration requires a sufficiently low reheating
temperature, the required reheating history moves the predicted tilt away
from the ACT-preferred value rather than toward it. This conclusion is not
universal: it can be avoided if the PQ sector couples only weakly to the
thermal bath, if its critical temperature is sufficiently high, or if the
post-inflationary equation of state differs from the matter-like case.

Within this thermally coupled minimal setup, viability requires three
simultaneous consistency conditions:
\begin{equation}
  \begin{aligned}
    \text{axion isocurvature:}\quad
    & \HI\ \text{small}
    &&\Longrightarrow\quad
    \epsilon_V\ll1,
    \\
    \text{observed red tilt:}\quad
    & \eta_V
      \simeq
      \frac{n_s-1}{2}
    &&=
      \mathcal O(-10^{-2}),
    \\
    \text{PQ non-restoration:}\quad
    & T_{\max}\lesssim T_{\rm PQ}
    &&\Longrightarrow\quad
    N_*\ \text{decreases}.
  \end{aligned}
  \label{eq:trilemma}
\end{equation}
The first two conditions require an exceptionally small potential slope
together with percent-level negative curvature. When the third condition
requires delayed matter-like reheating, it shortens the observable
inflationary interval and makes the predicted scalar spectrum still redder.

Within the uncorrelated, scale-invariant CDI analysis adopted here, including
ACT data changes the direct isocurvature-amplitude limit only modestly. In
fact, the P-ACT constraint is mildly weaker than the Planck-only result
(Sec.~\ref{subsec:benchmark})~\cite{Petretti:2026ayw}. The main impact of ACT
on the inflationary interpretation is instead its preference, in the data
combination considered above, for a larger value of $n_s$. The plateau and
hilltop results may be written schematically as
\begin{equation}
  1-n_s
  \simeq
  \frac{c}{N_*},
  \qquad
  c=
  \begin{cases}
    2,
    & \text{plateau},\\[3pt]
    \displaystyle\frac{2(p-1)}{p-2}>2,
    & \text{finite-$p$ hilltop},
  \end{cases}
  \label{eq:nsuniversality}
\end{equation}
reflecting their respective large-$N_*$ universality classes~\cite{Roest:2013fha}. By contrast, reproducing
$n_s\simeq0.9743$ at $N_*\simeq50$ would require $c_{\rm eff} \equiv N_*(1-n_s) \simeq 1.3$, below the coefficients of the simplest plateau and finite-power hilltop models.

The ACT-preferred tilt therefore motivates constructions in which the local
curvature is not fixed by these standard large-$N_*$ relations. Candidate
starting points include vacuum-dominated hybrid models, controlled
inflection-point potentials, and attractor models deformed away from their
asymptotic regime~\cite{Linde:1993cn,Dvali:1994ms,Allahverdi:2006iq,
McDonald:2025tfp,McDonald:2026pbf}. Such models must nevertheless reproduce
the extreme slope--curvature hierarchy derived in
Sec.~\ref{subsec:slopecurvature}, while remaining compatible with reheating
and PQ non-restoration. This is the precise sense in which ACT can leave the
direct axion-isocurvature ceiling nearly unchanged while substantially
sharpening the inflationary model-building problem.

\section{Which inflationary models remain viable?}
\label{sec:models}

\subsection{Two structural tests}
\label{subsec:discriminants}
Comprehensive surveys such as the
\emph{Encyclop{\ae}dia Inflationaris} commonly organize canonical
single-field inflationary models through their predictions in the
$(n_s,r)$ plane, including the dependence of those predictions on reheating~\cite{Martin:2013tda}. In the minimal pre-inflationary axion scenario,
however, the tensor amplitude is extraordinarily small. Using
Eq.~\eqref{eq:rH}, the order-one angle ceiling
$\HI<2.2\times10^7\GeV$ corresponds to
$r\lesssim7.9\times10^{-15}$, while the most permissive benchmark ceiling
$\HI<1.25\times10^{10}\GeV$ gives
$r\lesssim2.5\times10^{-9}$. Still smaller values remain allowed. These
amplitudes lie far below the sensitivity of current and planned CMB
polarization experiments~\cite{BICEPKeck:2024stm,LiteBIRD:2022cnt}.
The usual comparison of inflationary models in the $(n_s,r)$ plane therefore
becomes effectively one-dimensional: within the axion-compatible region,
$r$ is observationally indistinguishable from zero and ceases to provide a
useful discriminator. The relevant tests are instead structural.

The first is the \emph{scale test}: can the model realize
$\HI\lesssim10^{10}\GeV$, and preferably $\HI\sim10^7\GeV$, while reproducing
the observed scalar amplitude $\As$? A fixed-shape potential cannot generally
be moved into this regime by the rescaling
$V(\phi)\to cV(\phi)$ alone. At fixed $N_*$, such a rescaling leaves
$V'/V$, $V''/V$, and the potential slow-roll parameters unchanged, while
both $H_I^2$ and $\As$ scale proportionally to $c$. Lowering the normalization
therefore lowers the predicted scalar amplitude along with the inflationary
scale. Passing the scale test requires independently adjustable control of
the vacuum energy and the potential derivatives, a shape parameter that
suppresses $\epsilon_V$, or a departure from the canonical single-field
relation among $\As$, $\HI$, and $\epsilon_H$.

The second is the \emph{tilt test}: can the model reproduce an acceptable
$n_s$ at the reduced $e$-fold count implied by low-scale inflation? The
instantaneous reheating estimate derived in Sec.~\ref{sec:reheating} is
$\Ns^{\rm inst}\simeq48$--$52$, with smaller values possible when PQ
non-restoration requires delayed reheating with $w_{\rm reh}<1/3$. The
decisive question is whether the local curvature $\eta_V$ can be adjusted
independently or is locked, in the relevant asymptotic regime, to a relation
of the form
\begin{equation}
  n_s
  \simeq
  1-\frac{c}{\Ns},
  \qquad
  c\geq2.
  \label{eq:universalTiltModels}
\end{equation}
The simplest plateau models have $c=2$, while finite-power hilltop models
with $p>2$ have
$c=2(p-1)/(p-2)>2$~\cite{Roest:2013fha,Martin:2013tda}.
These models therefore predict relatively red spectra at the low values of
$\Ns$ relevant here.

Inverting Eq.~\eqref{eq:universalTiltModels} makes the tension transparent. A $c=2$ plateau reproduces the Planck 2018 central value $n_s=0.9649$ only for $\Ns\simeq57$, and the P-ACT-LB central value $n_s=0.9743$ only for $\Ns\simeq78$~\cite{Planck:2018jri,AtacamaCosmologyTelescope:2025blo}.
Both values exceed the instantaneous reheating range
$\Ns^{\rm inst}\simeq48$--$52$ obtained for the low inflationary scales
considered here. Over this range, the leading plateau relation predicts $n_s \simeq 0.9583\text{--}0.9615$.
If protecting the PQ symmetry requires delayed reheating with
$w_{\rm reh}<1/3$, the resulting decrease in $\Ns$ shifts the prediction
toward still smaller $n_s$. Because the P-ACT-LB central value is less red
than the Planck value, it sharpens the tilt test for universal plateau and
finite-power hilltop models. Conversely, it makes model classes with
independently adjustable curvature, including nearly scale-invariant
vacuum-dominated hybrid regimes, comparatively less disfavored~\cite{Linde:1993cn,Dvali:1994ms}.

A complete model must additionally sustain the required duration of
inflation, provide a graceful exit, reheat without restoring the PQ symmetry,
and satisfy the local hierarchy
\begin{equation}
  \frac{V_*^{1/4}}{\Mpl}
  \ll
  1,
  \qquad
  \left|
    \Mpl\frac{V_*'}{V_*}
  \right|
  \ll
  \left|
    \Mpl^2\frac{V_*''}{V_*}
  \right|
  \sim
  10^{-2},
  \label{eq:structuralreq}
\end{equation}
without violating observational bounds on running, features, or
non-Gaussianity, and without entering a diffusion-dominated regime along the
relevant inflationary trajectory.

To make the hierarchy explicit, fixing $\As$ at
$\HI\sim10^7\GeV$ requires $\epsilon_H \simeq \epsilon_V \sim 10^{-16}$
in canonical single-field slow-roll inflation, whereas the observed red tilt
requires $|\eta_V|\sim10^{-2}$. The dimensionless potential slope and
curvature are consequently separated by
\begin{equation}
  \frac{
    \left|\Mpl V_*'/V_*\right|
  }{
    \left|\Mpl^2V_*''/V_*\right|
  }
  =
  \frac{\sqrt{2\epsilon_V}}{|\eta_V|}
  \sim
  10^{-6}
  \label{eq:slopecurvatureRatio}
\end{equation}
at this scale, while the corresponding ratio of slow-roll parameters is
\begin{equation}
  \frac{\epsilon_V}{|\eta_V|}
  \sim
  10^{-14}.
\end{equation}
This hierarchy is illustrated in
Fig.~\ref{fig:derivativehierarchy}.

Within canonical single-field slow-roll inflation in which the inflaton
generates the observed curvature perturbation, passing the scale test requires
this hierarchy phenomenologically. A compelling ultraviolet realization must
also explain why it remains stable against radiative and Planck-suppressed
corrections, closely related to the usual eta problem~\cite{Copeland:1994vg,Dine:1995uk}. 
We now assess representative model classes against these requirements. 
The main conclusions are summarized in
Table~\ref{tab:modelsummary}.

\subsection{Models that fail the scale test}
\label{subsec:failscale}
In models whose dimensionless potential shape fixes the pivot-scale value of
$\epsilon_V$ once $N_*$ is specified, the measured scalar amplitude fixes the
inflationary Hubble scale. Such theories cannot be moved into the
axion-compatible window by lowering only the overall normalization of the
potential. At fixed $N_*$, the rescaling $V(\phi)\to cV(\phi)$
leaves the slow-roll trajectory unchanged but reduces $H_I^2$ and $\As$ by the same factor.

\paragraph{Large-field monomials.}
For the canonical potential
\begin{equation}
  V(\phi)
  =
  \lambda_p\phi^p,
  \qquad
  p>0,
\end{equation}
the leading large-$N_*$ slow-roll predictions are~\cite{Linde:1983gd,Martin:2013tda}
\begin{equation}
  n_s
  \simeq
  1-\frac{p+2}{2\Ns},
  \qquad
  r
  \simeq
  \frac{4p}{\Ns}.
  \label{eq:monomial}
\end{equation}
Quadratic inflation, for example, predicts
$r\simeq0.154\text{--}0.167$ for $\Ns\simeq48$--$52$. Using the measured
scalar amplitude, this corresponds to $\HI \simeq (9.7\text{--}10.1)\times10^{13}\GeV$.
It therefore exceeds the axion-isocurvature ceiling by several orders of
magnitude and is independently excluded by the BK18 limit
$r_{0.05}<0.036$ at 95\% confidence~\cite{BICEP:2021xfz}.

More generally, conventional positive-power monomials with
$p=\mathcal O(1)$ fail the scale test because
\begin{equation}
  \epsilon_V
  \simeq
  \frac{p}{4\Ns}
  =
  \mathcal O\!\left(\frac{1}{\Ns}\right),
\end{equation}
rather than the $\epsilon_V\sim10^{-16}$ required at
$\HI\sim10^7\GeV$. Their tilt depends on $p$ and need not always lie below
$0.96$, but varying $p$ within the conventional monomial regime cannot
suppress the tensor amplitude by the many orders of magnitude required here.

\paragraph{Starobinsky and standard metric Higgs inflation.}
Starobinsky $R^2$ inflation and metric Higgs inflation in its standard
large nonminimal-coupling attractor regime share the leading predictions~\cite{Starobinsky:1980te,Bezrukov:2007ep, Ellis:2013nxa}
\begin{equation}
  n_s
  \simeq
  1-\frac{2}{\Ns},
  \qquad
  r
  \simeq
  \frac{12}{\Ns^2}.
  \label{eq:starobinsky}
\end{equation}
Evaluating these relations at $\Ns\simeq48$--$52$ gives
\begin{equation}
  r
  \simeq
  (4.44\text{--}5.21)\times10^{-3},
  \qquad
  \HI
  \simeq
  (1.65\text{--}1.79)\times10^{13}\GeV
\end{equation}
after imposing the observed scalar normalization. Their self-consistent
high-scale reheating histories may give a somewhat different value of
$\Ns$, but this changes neither the order of magnitude of $\HI$ nor the
conclusion. The standard attractor regimes are incompatible with the minimal
pre-inflationary QCD axion dark-matter scenario considered here.

Over the same illustrative $e$-fold interval, Eq.~\eqref{eq:starobinsky}
gives
$n_s\simeq0.9583$--$0.9615$, below the P-ACT-LB central value and toward the
lower side of the Planck-preferred region~\cite{Planck:2018jri,AtacamaCosmologyTelescope:2025blo}. These models
therefore fail the scale test independently of this additional tilt tension.
This conclusion applies only to the standard metric attractor regime;
Higgs-inflation realizations that do not obey Eq.~\eqref{eq:starobinsky}
must be assessed separately.

\paragraph{The standard high fibre regime of fibre inflation.}
Fibre inflation provides an ultraviolet realization of plateau inflation in
type-IIB large volume compactifications. In the standard high fibre regime,
the potential satisfies the approximate relation~\cite{Cicoli:2008gp}
\begin{equation}
  r
  \simeq
  6(1-n_s)^2.
  \label{eq:fibreRelation}
\end{equation}
The representative reheating benchmark of Ref.~\cite{Cicoli:2008gp},
$N_e\simeq58$, gives
\begin{equation}
  n_s\simeq0.970,
  \qquad
  r\simeq0.005,
  \qquad
  \HI\simeq1.8\times10^{13}\GeV,
\end{equation}
where the last value follows after imposing the observed scalar
normalization. In the broad high fibre class described in that reference,
the adjustable parameters enter the leading inflationary potential through
its overall coefficient and do not independently suppress the slow-roll
parameters. The standard high fibre regime therefore fails the scale test.

This conclusion is not a no-go theorem for all string-motivated plateau
constructions. Compactification-dependent corrections, additional moduli,
alternative stabilization mechanisms, and specially tuned regions can modify
both the potential shape and its normalization~\cite{Baumann:2014nda}. Such realizations must be assessed individually rather than inferred from the standard fibre inflation benchmark.

\subsection{Models that can pass the scale test but face the tilt test}
\label{subsec:failtilt}
Some potentials can generate an extremely small $\epsilon_V$, and hence a
sufficiently small $r$ and $\HI$, while retaining a scalar tilt that is too
red at the reduced $e$-fold count relevant to the axion-compatible histories
considered here. In $\alpha$-attractors, a parameter suppresses $r$ without
changing the leading asymptotic prediction for $n_s$. In hilltop models,
including canonical natural inflation, the low scale instead arises because
the pivot lies very close to a local maximum, but the associated negative
curvature drives the spectrum toward small $n_s$. The tension is particularly
pronounced relative to the ACT DR6 central value, although its severity
depends on the data combination and on subleading corrections to the
asymptotic predictions~\cite{Planck:2018jri,AtacamaCosmologyTelescope:2025blo}.

\paragraph{$\alpha$-attractors.}
In the universal attractor regime, the leading large-$N_*$ predictions
are~\cite{Ellis:2013nxa, Kallosh:2013maa, Kallosh:2013yoa, Ellis:2019bmm, Kallosh:2015zsa, Carrasco:2015rva, Galante:2014ifa}
\begin{equation}
  n_s
  \simeq
  1-\frac{2}{\Ns},
  \qquad
  r
  \simeq
  \frac{12\alpha}{\Ns^2}.
  \label{eq:alphaattractor}
\end{equation}
Combining the second relation with Eq.~\eqref{eq:rH} gives
\begin{equation}
  \alpha
  \simeq
  3.4\times10^{-13}
  \left(
    \frac{\Ns}{50}
  \right)^2
  \left(
    \frac{\HI}{10^7\GeV}
  \right)^2.
  \label{eq:alphaTiny}
\end{equation}
Thus, the inflationary scale can formally be lowered into the axion-compatible
window by taking an extremely small pole residue. This suppresses $r$ without
changing the leading attractor prediction for the tilt, which remains
\begin{equation}
  n_s
  \simeq
  0.9583\text{--}0.9615
  \qquad
  \text{for}
  \qquad
  \Ns\simeq48\text{--}52.
\end{equation}
The universal attractor regime therefore lies below the ACT DR6 central
value at the relevant $\Ns$, and delayed reheating introduced to protect the
PQ symmetry lowers $\Ns$ and reddens the prediction further.

In supergravity realizations, $\alpha$ is inversely related to the curvature
of the inflaton K\"ahler manifold. Taking the extremely small value required
by Eq.~\eqref{eq:alphaTiny} is therefore a phenomenological possibility, not
by itself a demonstration of a controlled ultraviolet construction. This
conclusion applies to the universal asymptotic regime rather than to every
model bearing the $\alpha$-attractor label. Subleading terms, localized
features, uplifting, or departure from the universal attractor regime can modify
$n_s$ and potentially raise it. 
The result then depends on the detailed
potential rather than on the universal relation in
Eq.~\eqref{eq:alphaattractor}. Structurally, such realizations belong to the
independent-curvature class discussed below. See Refs.~\cite{Ellis:2021kad, Ellis:2025zrf} 
for a more extensive discussion on $\alpha$-attractors.

\paragraph{Canonical natural inflation.}
For the original single cosine potential
\begin{equation}
  V(\phi)
  =
  \Lambda^4
  \left[
    1-\cos\!\left(\frac{\phi}{f}\right)
  \right],
  \label{eq:naturalInflationPotential}
\end{equation}
where $f$ is the inflaton decay constant and is unrelated to the QCD axion
decay constant $\fa$, the parameter $f/\Mpl$ controls both the slope and
curvature of the potential~\cite{Freese:1990rb,Martin:2013tda}. Defining
\begin{equation}
  u_*
  \equiv
  \cos^2\!\left(\frac{\phi_*}{2f}\right)
  =
  \frac{2f^2}{2f^2+\Mpl^2}
  \exp\!\left(
    -\frac{\Ns\Mpl^2}{f^2}
  \right),
  \label{eq:naturalInflationU}
\end{equation}
the leading slow-roll predictions can be written as
\begin{equation}
  n_s
  \simeq
  1-
  \frac{\Mpl^2}{f^2}
  \frac{1+u_*}{1-u_*},
  \qquad
  r
  \simeq
  8\frac{\Mpl^2}{f^2}
  \frac{u_*}{1-u_*}.
  \label{eq:naturalInflationObservables}
\end{equation}

In the small-$f$ hilltop regime, $u_*$ is exponentially suppressed. The
tensor amplitude and inflationary scale can consequently be extremely small,
but
\begin{equation}
  n_s
  \longrightarrow
  1-\frac{\Mpl^2}{f^2},
  \qquad
  u_*\ll1,
  \label{eq:naturalInflationSmallF}
\end{equation}
so the spectrum becomes unacceptably red. For example, the leading slow-roll
expressions give
\begin{equation}
  f=1.5\Mpl,
  \qquad
  \Ns=48\text{--}52:
  \qquad
  r\simeq
  2.7\times10^{-10}\text{--}1.6\times10^{-9},
  \qquad
  \HI\simeq
  (4.1\text{--}9.9)\times10^9\GeV,
\end{equation}
but simultaneously give $n_s\simeq0.556$. This illustrates that canonical
natural inflation can pass the scale test in a finely positioned hilltop
regime, but only at the cost of a grossly unacceptable tilt.

Increasing $f$ raises $n_s$ toward the quadratic-inflation limit, but also
raises $r$ and $\HI$. In the region with a phenomenologically acceptable
tilt, the tensor amplitude is of order $10^{-2}$--$10^{-1}$ and the
inflationary scale is of order $10^{13}$--$10^{14}\GeV$. The original
single cosine model therefore has no parameter region that simultaneously passes the scale and tilt tests~\cite{Planck:2018jri,BICEP:2021xfz}. Hybrid natural inflation models evade this conclusion by introducing an independent vacuum-energy contribution and a separate waterfall sector. They belong to the independent-curvature class discussed below.

\paragraph{Pure hilltop and Coleman--Weinberg new inflation.}
Consider a small-field hilltop potential whose leading departure from the
maximum is~\cite{Linde:1981mu, Albrecht:1982wi, Barenboim:2013wra, 
Boubekeur:2005zm, Kumekawa:1994gx}
\begin{equation}
  V(\phi)
  \simeq
  V_0
  \left[
    1-
    \left(
      \frac{\phi}{\mu}
    \right)^p
    +\cdots
  \right],
  \qquad
  p>2.
  \label{eq:hilltopPotential}
\end{equation}
The shape parameter $\mu/\Mpl$ and the proximity of the pivot to the maximum
can make $\epsilon_V$ extremely small. The scalar normalization then fixes
$V_0$ at a correspondingly low value, allowing these models to pass the scale
test. In the asymptotic small-field regime, however,
the leading tilt is
\begin{equation}
  n_s
  \simeq
  1-
  \frac{2(p-1)}
       {(p-2)\Ns},
  \qquad
  p>2.
  \label{eq:hilltopModelTilt}
\end{equation}
Since
\begin{equation}
  \frac{2(p-1)}{p-2}
  =
  2+\frac{2}{p-2}
  >
  2,
\end{equation}
the asymptotic hilltop prediction is always redder than
$n_s\simeq1-2/\Ns$ at fixed $\Ns$, approaching the plateau result only in the
formal limit $p\to\infty$. For the commonly studied quartic hilltop,
\begin{equation}
  p=4:
  \qquad
  n_s
  \simeq
  1-\frac{3}{\Ns}
  \simeq
  0.9375\text{--}0.9423
  \qquad
  \text{for}
  \qquad
  \Ns\simeq48\text{--}52,
\end{equation}
which is substantially too red.

The small-field Coleman--Weinberg potential~\cite{Coleman:1973jx,
Barenboim:2013wra},
\begin{equation}
  V_{\rm CW}(\phi)
  =
  A\phi^4
  \left[
    \ln\!\left(\frac{\phi}{v}\right)
    -\frac{1}{4}
  \right]
  +
  \frac{Av^4}{4},
  \label{eq:ColemanWeinbergPotential}
\end{equation}
has a quartic hilltop dressed by a logarithm. In its asymptotic
$\phi\ll v$ regime, it approaches
$n_s\simeq1-3/\Ns$ up to logarithmic corrections and therefore inherits the
quartic-hilltop tilt problem~\cite{Barenboim:2013wra}. Delayed reheating
introduced to prevent PQ restoration reduces $\Ns$ and reddens both the pure
hilltop and small-field Coleman--Weinberg predictions further.

Quadratic hilltops, Coleman--Weinberg models away from their asymptotic
small-field limit, and hilltop potentials supplemented by additional
operators can possess greater freedom than
Eq.~\eqref{eq:hilltopModelTilt} suggests. Viable low-scale regions may then
exist if the pivot lies outside the universal asymptotic regime or if
additional terms independently modify $\eta_V$. Such constructions no longer
derive their viability from the pure hilltop form alone: structurally, they
have acquired the independent curvature control characteristic of the model
classes considered next.

\subsection{Models with sufficient freedom to pass both tests}
\label{subsec:independent}
For canonical single-field slow-roll evolution in which the inflaton
generates the observed curvature perturbation, the negligible tensor
amplitude also implies very little field motion across the observable CMB
window. Using the most permissive minimal-benchmark ceiling,
$r\lesssim2.5\times10^{-9}$, and assuming that $r$ varies negligibly across
$\Delta N_{\rm CMB}\simeq8$, the local Lyth relation gives
\begin{equation}
  \frac{|\Delta\phi_{\rm CMB}|}{\Mpl}
  \simeq
  \Delta N_{\rm CMB}
  \sqrt{\frac{r}{8}}
  \lesssim
  1.4\times10^{-4}
  \left(
    \frac{\Delta N_{\rm CMB}}{8}
  \right).
  \label{eq:localLythModels}
\end{equation}
This is a local statement: it constrains the field motion while observable
CMB modes leave the horizon, not the total excursion from the onset to the
end of inflation.

Models capable of satisfying both the scale and tilt tests generally share a
further structural property: the vacuum energy, local slope, local curvature,
and termination of inflation are controlled by at least partially independent
ingredients. This separation permits a low value of $\HI$ together with
$\epsilon_V\ll|\eta_V|$, while allowing $n_s$ to be adjusted at
$\Ns\simeq50$ rather than being locked to a universal large-$\Ns$ relation.
Such flexibility is not automatic and is generally obtained at the cost of
parameter tuning, radiative sensitivity, additional fields, or a restricted
inflationary trajectory. The examples below are representative rather than
exhaustive. A broader catalogue of inflationary potentials is given in
Ref.~\cite{Martin:2013tda}.

\paragraph{Vacuum-dominated hybrid inflation.}
Hybrid inflation~\cite{Linde:1993cn} provides the clearest realization of
this structure. Along the inflationary valley, the effective potential takes
the form
\begin{equation}
  V(\phi)
  =
  V_0+\Delta V(\phi),
  \qquad
  |\Delta V(\phi)|\ll V_0,
  \label{eq:vacdom}
\end{equation}
so that the nearly constant vacuum energy determines $\HI$, whereas
$\Delta V$ controls the local slope and curvature. Inflation terminates when
a second, waterfall field becomes unstable rather than when slow roll fails.
Consequently, the vacuum scale, pivot-scale derivatives, duration of
inflation, and end point need not be determined by a single potential shape.
Although the observable phase can be effectively single-field, the complete
theory is intrinsically multifield.

For supersymmetric F-term hybrid inflation, the standard superpotential is
\begin{equation}
  W
  =
  \kappa S
  \left(
    \Phi\bar\Phi-M^2
  \right),
  \label{eq:Fterm}
\end{equation}
and, along the inflationary valley,
\begin{equation}
  V_0
  \simeq
  \kappa^2M^4,
  \qquad
  \HI
  \simeq
  \frac{\kappa M^2}
       {\sqrt{3}\Mpl}
  \label{eq:hybridH}
\end{equation}
up to supergravity and radiative corrections~\cite{Dvali:1994ms}. The
parameters $\kappa$ and $M$ determine the vacuum scale, while radiative
corrections, supersymmetry-breaking terms, and nonminimal K\"ahler operators
control the slope and curvature. These ingredients are not fully independent,
but their separation is substantially greater than in a fixed-shape
single-field potential.

Minimal radiatively driven F-term hybrid inflation typically predicts a
nearly scale-invariant spectrum with $n_s\gtrsim0.98$, and can become blue
when supergravity corrections are important. Soft supersymmetry-breaking
terms or nonminimal K\"ahler operators can instead generate the required
negative curvature~\cite{Bastero-Gil:2006zpr}. For a conventional
quartic K\"ahler correction, the induced contribution to $\eta_V$ is
parametrically of the same order as its dimensionless coefficient, so the
required curvature corresponds to a percent-level coefficient. Protecting
this coefficient from generic supergravity corrections remains a controlled
version of the eta problem.

Supersymmetric D-term hybrid inflation provides a related vacuum-dominated
construction~\cite{Binetruy:1996xj,Halyo:1996pp}. An explicit analysis of
low-scale F- and D-term hybrid inflation with axion dark matter finds that
avoiding excessive axion isocurvature requires Yukawa or gauge couplings of
order $10^{-3}$ or smaller, with additional model-dependent restrictions
from supersymmetry breaking and cosmic strings~\cite{Schmitz:2018nhb}.
Smooth and shifted hybrid variants modify the inflationary trajectory and can
avoid post-inflationary production of the corresponding topological defects
because the relevant symmetry is already broken during inflation~\cite{Lazarides:1995vr, Jeannerot:2002wt}.

The virtue of hybrid inflation is therefore its parameter structure, not an
automatic guarantee of viability. Waterfall fluctuations and multifield
conversion must be checked explicitly~\cite{Lyth:2010zq}, together with
defect production, radiative stability, reheating, and PQ non-restoration.

\paragraph{Running-mass inflation.}
Running-mass inflation uses radiative evolution to separate the local slope
from the curvature. A representative potential is
\begin{equation}
  V(\phi)
  =
  V_0
  +
  \frac{1}{2}m^2(\phi)\phi^2
  +\cdots,
  \qquad
  \frac{\dd m^2}{\dd\ln\phi}
  \neq
  0.
  \label{eq:runningmass}
\end{equation}
Renormalization-group evolution can produce a cancellation in the first
derivative near the pivot while leaving a percent-level second derivative,
thereby realizing $\epsilon_V\ll|\eta_V|$
~\cite{Stewart:1996ey, Covi:1998mb, Covi:1998jp}. The same mechanism generally makes $\eta_V$ scale
dependent and can therefore generate appreciable running or higher scale
dependence. Running-mass inflation thus correlates the tilt and its running
rather than providing arbitrary independent tunability. The observational
consistency of $\alpha_s$ with zero constrains both the magnitude and the
variation of the radiative correction across the CMB window~\cite{Planck:2018jri,AtacamaCosmologyTelescope:2025nti}. Its viability also
depends on whether the small pivot-scale slope and the required running remain
stable under ultraviolet threshold corrections.

\paragraph{Inflection point and MSSM inflation.}
Choose a reference point $\phi_0$ near a quasi-inflection point and expand the
potential as
\begin{equation}
  V(\phi)
  =
  V_0
  +
  a\,\Delta\phi
  +
  \frac{b}{2}\,\Delta\phi^2
  +
  \frac{c}{3!}\,\Delta\phi^3
  +\cdots,
  \qquad
  \Delta\phi
  \equiv
  \phi-\phi_0,
  \qquad
  V_0\equiv V(\phi_0),
  \label{eq:inflection}
\end{equation}
where $a=V'(\phi_0)$, $b=V''(\phi_0)$, and
$c=V'''(\phi_0)$. A mathematical inflection point has $b=0$, whereas a
stationary inflection point, often called a saddle point in this context,
additionally has $a=0$. In a quasi-inflection construction, $a$ and $b$ are
small but nonzero. At the pivot,
\begin{equation}
  \begin{aligned}
    V_*'
    &=
    a+b\,\Delta\phi_*
    +\frac{c}{2}\,\Delta\phi_*^2+\cdots,
    \\
    V_*''
    &=
    b+c\,\Delta\phi_*+\cdots.
  \end{aligned}
  \label{eq:inflectionDerivatives}
\end{equation}
The pivot can therefore lie on the negative-curvature side of the
quasi-inflection point, with $V_*''<0$, while the coefficients conspire to
keep $|V_*'|$ extremely small. The higher derivatives control how rapidly
the curvature changes, the duration of the near-inflectionary phase, and the
scale dependence of the spectrum. If the underlying theory provides
sufficiently independent control of these coefficients and of $V_0$, the
inflationary scale is not fixed by a universal potential shape~\cite{Hotchkiss:2011am}.

MSSM flat-direction inflation provides a particle-physics realization of
this structure. The inflaton parameterizes a gauge-invariant $D$-flat
combination of MSSM fields, such as the $LLe$ or $udd$ directions, while soft
masses, trilinear $A$-terms, and nonrenormalizable operators generate a
near-saddle or quasi-inflection point. For the standard dimension-six
lifting operator, the exact saddle-point condition is
$A^2=40m_\phi^2$, with a small departure from this relation generating the
residual slope required for slow roll~\cite{Allahverdi:2006iq}. The classic realization operates at $\HI\sim\mathcal O(1\text{--}10)\GeV$, demonstrating that
inflection-point inflation is not intrinsically tied to a high inflationary
scale. Its principal costs are the sensitive relation among the potential
parameters, its required radiative stability, and potentially restrictive
initial conditions. In particular completions, the initial-condition problem
can be ameliorated by an earlier false-vacuum inflationary phase that attracts
the field toward the inflectionary region~\cite{Allahverdi:2008bt}.

Near the flattest part of the potential, a useful diagnostic for a light
canonical inflaton in the slow-roll regime is to compare the classical
displacement during one Hubble time with the typical quantum fluctuation:
\begin{equation}
  \delta\phi_{\rm cl}
  \equiv
  \frac{|\dot\phi|}{H}
  \simeq
  \frac{|V'|}{3H^2}
  \gtrsim
  \delta\phi_{\rm q}
  \equiv
  \frac{H}{2\pi}.
  \label{eq:diffusion}
\end{equation}
If this condition is marginal or violated, quantum diffusion cannot be
neglected and a stochastic treatment is required~\cite{Starobinsky:1986fx}. Moreover, a sufficiently flat
quasi-inflectionary region can generate a non-attractor or ultra-slow-roll
phase, during which the curvature perturbation evolves outside the horizon
and the usual horizon-crossing slow-roll expressions need not apply~\cite{Kinney:2005vj, Germani:2017bcs}. Any resulting enhancement of the
small-scale power spectrum, primordial-black-hole abundance, or
non-Gaussian statistics must then be calculated from the full background and
perturbation evolution, supplemented by stochastic methods where necessary~\cite{Germani:2017bcs, Ezquiaga:2018gbw}.

Among the classes considered here, vacuum-dominated hybrid inflation provides
the clearest structural route to a low inflationary scale because the vacuum
energy, inflaton derivatives, and end of inflation arise from partly
independent ingredients. Running-mass and inflection-point models provide
effectively single-field alternatives, but require direct control of
radiative corrections or finely related potential coefficients. None of
these structures automatically guarantees viability: the scalar amplitude,
tilt, running, exit, reheating history, and PQ non-restoration must still be
verified in a complete model.

\paragraph{Small-field hilltop inflation.}
An explicit single-field realization of a very low inflationary scale is
provided by the potential~\cite{Kumekawa:1994gx, Ema:2017rkk}
\begin{equation}
  V(\phi)
  =
  \Lambda^4
  \left[
    1-
    \left(
      \frac{\phi}{v_\phi}
    \right)^n
  \right]^2,
  \qquad
  n>2,
  \qquad
  v_\phi\ll\Mpl \,.
  \label{eq:lowScaleHilltop}
\end{equation}
At leading order,
\begin{equation}
  n_s
  \simeq
  1-
  \frac{2(n-1)}
       {(n-2)N_*},
  \qquad
  r
  \simeq
  \frac{16n}{N_*(n-2)}
  \left[
    \frac{1}{2N_*n(n-2)}
    \left(
      \frac{v_\phi}{\Mpl}
    \right)^2
  \right]^{\frac{n}{n-2}}.
  \label{eq:lowScaleHilltopObservables}
\end{equation}
The tensor amplitude, and hence the inflationary Hubble scale, can be made
extremely small by taking $v_\phi/\Mpl\ll1$. The model therefore provides a
concrete example that passes the scale test without requiring a separate
waterfall field. Its leading prediction for the scalar tilt, however, is the
usual $p>2$ hilltop result and is too red for the reduced $e$-fold range
considered here. For instance, $n=6$ and $N_*=50$ give
$n_s\simeq0.950$.

A small Planck-suppressed quadratic operator can modify the pivot-scale
curvature and improve the tilt~\cite{Ema:2017rkk}, but this constitutes an additional deformation of the minimal hilltop potential. Such a model belongs structurally to the independent-curvature class rather than to the pure hilltop class.

\subsection{Beyond canonical cold single-field inflation}
\label{subsec:beyond}
The two structural tests above assume that the observed curvature perturbation
is generated by a canonical, cold, slowly rolling inflaton. Relaxing this
assumption opens additional possibilities, but does not by itself suppress
the axion fluctuation generated during inflation. Provided the axion remains
a light, canonically normalized spectator with the transfer and relic history
assumed in Sec.~\ref{sec:axionbound}, the direct isocurvature bound on $\HI$
continues to apply.\footnote{A short late period of thermal inflation, driven by a flaton trapped near the origin by finite-temperature effects~\cite{Lyth:1995ka}, is sometimes
used to dilute unwanted relics. A conventional thermal inflation period does
not erase an axion fluctuation that was already generated on superhorizon
scales, although it can dilute the axion abundance or modify its subsequent
transfer, depending on the cosmological chronology. Such effects belong to
the nonminimal abundance and transfer histories discussed in
Sec.~\ref{sec:loopholes}, rather than to the inflationary model discussion here.}

\paragraph{Spectator-generated curvature perturbations.}
Curvaton and modulated reheating mechanisms remove the identification of the
observed scalar amplitude with the inflaton contribution
\begin{equation}
  \PRR^{(\phi)}(k_*)
  \simeq
  \frac{\HI^2}
       {8\pi^2\epsilon_H\Mpl^2}.
  \label{eq:AsSpectatorComparison}
\end{equation}
The observed amplitude $\As$ can then be generated predominantly by another
degree of freedom, so the inflaton itself need not have an extraordinarily
small $\epsilon_H$ at fixed $\HI$~\cite{Lyth:2001nq, Dvali:2003em}. These mechanisms do not, however, remove the axion fluctuation entering Eq.~\eqref{eq:PIIexact}. The direct upper bound on $\HI$ remains unless the additional sector also modifies the inflationary axion normalization, its subsequent transfer, its relic abundance, or its correlation with the adiabatic mode.

Multi-source models must additionally satisfy constraints on residual and
correlated isocurvature perturbations and on local non-Gaussianity~\cite{Planck:2019kim}. The same logic applies to genuine multifield
inflation, in which entropic perturbations can be converted into curvature
perturbations along a turning trajectory~\cite{Gordon:2000hv}. Such conversion
changes the origin and normalization of $\As$, but it does not alter the
directly generated axion fluctuation unless the additional fields also couple
to, or otherwise modify, the axion sector.

\paragraph{Noncanonical kinetic terms.}
For a slowly varying single-clock $P(X,\phi)$ theory in Einstein gravity, the
leading scalar and tensor spectra give
\begin{equation}
  \As
  \simeq
  \frac{\HI^2}
       {8\pi^2\Mpl^2\epsilon_H c_s},
  \qquad
  r
  \simeq
  16\epsilon_H c_s,
  \label{eq:noncanonicalSpectra}
\end{equation}
where $c_s$ is the scalar sound speed~\cite{Garriga:1999vw,Chen:2006nt}. At fixed $\HI$ and $\As$, the product
$\epsilon_Hc_s$ is fixed, so a reduced sound speed permits a larger
$\epsilon_H$ than in the canonical theory. This can weaken the inference of
an extremely small Hubble slow-roll parameter, although the available freedom
is strongly constrained by equilateral non-Gaussianity and by the
strong-coupling scale of the kinetic effective theory~\cite{Chen:2006nt,Planck:2019kim}. The scalar tilt also depends on the
evolution of both $\epsilon_H$ and $c_s$, so it is not determined by a
canonical potential curvature alone.

For the standard tensor vacuum,
Eq.~\eqref{eq:noncanonicalSpectra} still implies
\begin{equation}
  r
  \simeq
  \frac{2\HI^2}
       {\pi^2\Mpl^2\As}.
  \label{eq:noncanonicalrH}
\end{equation}
Thus, changing $c_s$ does not alter the relation between $\HI$ and the tensor
amplitude at fixed observed $\As$. Moreover, the canonical relation between
$r$ and $\dd\phi/\dd N$ in Eq.~\eqref{eq:lythdiff} no longer follows:
the appropriate field-range statement depends on the kinetic normalization
and on the ultraviolet realization of the effective theory, and cannot in
general be expressed as a universal bound on the coordinate displacement
$\Delta\phi$~\cite{Baumann:2011ws}.

The direct axion bound on $\HI$ likewise remains unchanged provided the axion
is a light, canonically normalized spectator with the vacuum fluctuation
spectrum assumed in Sec.~\ref{sec:axionbound}. Noncanonical inflaton dynamics
can therefore modify the inferred background hierarchy and field excursion
without, by itself, evading the axion constraint.

\paragraph{Warm inflation.}
Dissipation and thermal fluctuations can enhance the scalar power spectrum,
allowing low-scale inflation without the extremely small value of
$\epsilon_V$ inferred from the cold canonical formula~\cite{Berera:1995ie, Berera:2008ar, Bastero-Gil:2016qru}. A consistent realization nevertheless requires a microscopic origin for the dissipative coefficient, control over thermal and radiative corrections to the inflaton potential, and a simultaneous treatment of the PQ sector.

If the PQ symmetry remains broken and the axion is sufficiently decoupled
from the radiation bath, the vacuum axion fluctuation and the corresponding
bound on $\HI$ remain essentially unchanged. If instead the axion
thermalizes, acquires a substantial thermal mass, or the PQ symmetry is
restored or broken during warm inflation, the minimal fluctuation and
chronology assumed above no longer apply. Indeed, the PQ transition can occur
during warm inflation in explicit constructions~\cite{Rosa:2021gbe}. The inflaton, radiation, radial PQ field, and axion must then be evolved within a common thermal framework. Warm inflation is therefore not merely a modification of the scalar normalization: it can also change the PQ history on which the pre-inflationary axion scenario depends.

\paragraph{Nonminimal couplings and Palatini gravity.}
Palatini Higgs inflation and related nonminimally coupled models can suppress
$r$ relative to their metric counterparts and realize an intermediate or low
inflationary scale. For example, Palatini Higgs inflation was proposed as a
way of obtaining $\HI\sim10^8\GeV$ while accommodating pre-inflationary axion
dark matter~\cite{Tenkanen:2019xzn}. Compatibility with the axion cannot,
however, be inferred from $\HI$ or $r$ alone. The canonical normalization,
effective cutoff, reheating dynamics, and inflationary axion radius must all
be evaluated in the Einstein frame theory.

A nonminimal coupling of the inflaton can reduce the effective angular radius
of the axion and strengthen the isocurvature constraint, whereas a suitable
nonminimal coupling of the radial PQ field can increase the inflationary
radius $\fI$. The competition between these effects, and the range in which
the radial coupling helps without significantly backreacting on inflation,
differs between the metric and Palatini formulations~\cite{Rigouzzo:2025ycb}. In the minimal Higgs-inflation setup analyzed in Ref.~\cite{Rigouzzo:2025hza}, the field-dependent axion normalization enhances rather than suppresses isocurvature, restricting the axion dark matter fraction to approximately $\gamma_a \lesssim 10^{-5}$. This result illustrates why a low value of $\HI$ is not, by itself, sufficient.
Nonminimal models must be analyzed using the complete Einstein-frame kinetic
sector, as discussed in Sec.~\ref{subsec:nonminimal}; neither universal
viability nor universal exclusion follows from the Palatini formulation
alone.

The resulting structural classification is summarized in
Table~\ref{tab:modelsummary}. The entries indicate whether a class possesses
the necessary freedom in principle; they do not replace a complete
model-by-model analysis.

\begin{table}[ht!]
\centering

\begingroup
\footnotesize
\setlength{\tabcolsep}{3.0pt}
\renewcommand{\arraystretch}{0.98}

\begin{tabularx}{\textwidth}{
  @{}
  >{\raggedright\arraybackslash}p{0.18\textwidth}
  >{\centering\arraybackslash}p{0.105\textwidth}
  >{\centering\arraybackslash}p{0.13\textwidth}
  >{\raggedright\arraybackslash}p{0.235\textwidth}
  >{\raggedright\arraybackslash}X
  @{}
}
\toprule
Class
&
Scale test status
&
Tilt at $\Ns=48$--$52$
&
Principal obstacle
&
Required structure or cost
\\
\midrule

\multicolumn{5}{@{}l}{
  \itshape Standard regimes that fail the scale test\/
}
\\[-2pt]
\cmidrule(lr){1-5}

Monomials
&
Fail
&
$1-(p+2)/(2\Ns)$
&
Tensor amplitude is far too large
&
A qualitatively different potential shape
\\

Starobinsky / metric Higgs
&
Fail
&
$0.958$--$0.962$
&
Normalization fixes $\HI\sim10^{13}\GeV$
&
A modified realization outside the standard attractor regime
\\

Standard string plateaus
&
Fail generically
&
Usually plateau-like
&
The normalization and compactification data typically fix a high scale
&
Additional hierarchies, sequestering, or tunable compactification data
\\

\addlinespace[1pt]
\multicolumn{5}{@{}l}{
  \itshape Models that can pass the scale test but face the tilt test\/
}
\\[-2pt]
\cmidrule(lr){1-5}

$\alpha$-attractors
&
Formal pass
&
$0.958$--$0.962$
&
Universal tilt lies below the ACT DR6 central value
&
An extremely small value of $\alpha$ or nonuniversal corrections
\\

Canonical natural inflation
&
Formal pass
&
Unacceptably red at low $\HI$
&
A sufficiently small tensor amplitude is correlated with a very red tilt
&
Additional harmonics, independent curvature control, or a hybrid exit
\\

Pure $p>2$ hilltop / Coleman--Weinberg
&
Pass
&
$1-c/\Ns$, $c>2$
&
The asymptotic tilt is redder than the plateau prediction
&
Additional operators or a nonasymptotic pivot region
\\

\addlinespace[1pt]
\multicolumn{5}{@{}l}{
  \itshape Canonical models with independent curvature control\/
}
\\[-2pt]
\cmidrule(lr){1-5}

\textbf{Vacuum-dominated hybrid}
&
\textbf{Can pass}
&
\textbf{Adjustable}
&
Minimal variants may be blue, waterfall dynamics and possible defects
&
Soft or K\"ahler corrections, controlled waterfall dynamics, and reheating
\\

\textbf{Running-mass inflation}
&
\textbf{Can pass}
&
\textbf{Adjustable}
&
Running, higher scale dependence, and radiative stability
&
Controlled RG evolution and ultraviolet thresholds
\\

\textbf{Inflection point / MSSM inflation}
&
\textbf{Can pass}
&
\textbf{Adjustable}
&
Coefficient tuning, features, or quantum diffusion
&
Controlled coefficients and suitable initial conditions
\\

\textbf{Quadratically corrected hilltop}
&
\textbf{Can pass}
&
\textbf{Adjustable}
&
Quadratic-term tuning, radiative stability, and initial conditions
&
A controlled mass term, higher-order completion, and consistent exit
\\

\addlinespace[1pt]
\multicolumn{5}{@{}l}{
  \itshape Beyond canonical cold single-field inflation\/
}
\\[-2pt]
\cmidrule(lr){1-5}

Curvaton / modulated reheating
&
Relaxes the $\As$ inference
&
Model dependent
&
Residual or correlated isocurvature and local $f_{\rm NL}$
&
An additional sector, the direct axion bound remains
\\

Noncanonical $P(X,\phi)$
&
Limited relief
&
Model dependent
&
Equilateral non-Gaussianity and strong coupling
&
A controlled kinetic EFT, the direct axion bound remains
\\

Warm inflation
&
Can pass
&
Model dependent
&
Microscopic dissipation, thermal corrections, and the PQ chronology
&
A complete and controlled thermal sector
\\

Nonminimal / Palatini models
&
Model dependent
&
Model dependent
&
Einstein-frame axion normalization, cutoff, and reheating
&
A complete Einstein frame kinetic and reheating analysis
\\

\bottomrule
\end{tabularx}
\endgroup

\caption{Structural assessment of representative inflationary classes in the
minimal pre-inflationary light axion scenario. ``Adjustable'' denotes
sufficient freedom, in principle, to modify the local curvature and reproduce
the observed tilt. Boldface marks canonical classes possessing the structural
freedom to pass both the scale and tilt tests in suitable regions. It does not
imply that every realization is viable. Reheating, PQ non-restoration,
radiative stability, and other model-specific constraints must still be
imposed. Scenarios in which the axion becomes heavy during inflation modify
the light axion premise and therefore lie outside this classification.}
\label{tab:modelsummary}
\end{table}

As summarized in Table~\ref{tab:modelsummary}, the audit yields a simple
structural conclusion. In the minimal pre-inflationary light axion scenario,
many widely studied fixed-shape and universal attractor models either fail the
scale test in their standard regimes or face the tilt test at the reduced
values of $\Ns$ relevant here. Among canonical constructions, quadratically corrected hilltop, vacuum-dominated hybrid, running-mass, and inflection point models can realize the required slope--curvature hierarchy. Vacuum-dominated hybrid inflation provides the clearest structural route because the vacuum energy, local inflaton derivatives, and end of inflation arise from partly independent ingredients. Quadratically corrected hilltop models provide a simpler single-field route, but require a radiatively stable curvature term, a
controlled higher-order completion, and suitable initial conditions.
Running-mass and inflection-point models provide further effectively
single-field alternatives, at the cost of controlled radiative evolution or
finely related potential coefficients.

This flexibility comes at the cost of tuning, ultraviolet sensitivity, or
additional dynamics rather than attractor universality. Each surviving class
also carries characteristic costs and potential diagnostics: waterfall
dynamics and possible defect production in hybrid models, running and higher
scale dependence in running-mass models, and features, initial-condition
sensitivity, or quantum diffusion near inflection points.

Relaxing the canonical cold single-field assumptions can change the inference
from $\As$ to the inflaton dynamics. It does not, however, remove the direct
axion bound on $\HI$ if the axion remains a light, canonically normalized
spectator with the standard transfer and relic history. Evading that bound
requires modifying at least one axion-side ingredient, i.e., its inflationary
normalization or mass, primordial spectrum, transfer or correlation,
thermal history, or relic abundance. The axion constraint therefore does
more than favor a lower inflationary scale: it requires either a specially
organized inflaton sector or a controlled departure from the assumptions of
the minimal axion scenario.

\section{How axion dynamics suppress or evade the bound}
\label{sec:loopholes}
Equation~\eqref{eq:masterboundPII} applies when the axion is light at CMB
horizon exit, its inflationary canonical normalization is known, its
perturbation survives until the QCD epoch, and its late abundance is
determined by the local misalignment angle. Nonminimal scenarios can therefore
be classified by which ingredient they modify: the fluctuation normalization,
the inflationary axion mass, the post-inflationary abundance and transfer, or
the PQ history and primordial initial condition. These possibilities are
physically distinct and need not lead to the same residual signatures.

\subsection{Changing the inflationary normalization}
\label{subsec:radial}

Write the complex PQ field as
\begin{equation}
  P
  =
  \frac{\rho}{\sqrt{2}}\,
  e^{i\vartheta},
  \qquad
  \theta
  =
  N_{\rm DW}\vartheta ,
  \label{eq:PQfield}
\end{equation}
where $\theta$ is the physical axion misalignment angle. We use the
convention $\rho_0 = N_{\rm DW}\fa$, with $\rho_0$ the late-time radial expectation value. For a canonical PQ kinetic term,
\begin{equation}
  \mathcal L_{\rm kin}
  =
  \frac{1}{2}(\partial\rho)^2
  +
  \frac{1}{2}\rho^2(\partial\vartheta)^2
  =
  \frac{1}{2}(\partial\rho)^2
  +
  \frac{1}{2}
  \left(
    \frac{\rho}{N_{\rm DW}}
  \right)^2
  (\partial\theta)^2 .
  \label{eq:PQkinetic}
\end{equation}
The canonical angular radius during inflation is therefore
\begin{equation}
  \fI
  =
  \frac{\rho_I}{N_{\rm DW}}.
\end{equation}
Provided that the angular mode is light and $\rho$ varies slowly during
horizon crossing, its fluctuation is
\begin{equation}
  \delta\theta_*
  \simeq
  \frac{\HI}{2\pi\fI}
  =
  \frac{N_{\rm DW}\HI}{2\pi\rho_I}.
  \label{eq:largeradius}
\end{equation}

An inflationary PQ potential, such as one containing a negative
Hubble-induced mass, can stabilize the radial field at
$\rho_I\gg\rho_0=N_{\rm DW}\fa$. Equivalently,
\begin{equation}
  \frac{\fI}{\fa}
  =
  \frac{\rho_I}{\rho_0}
  \gg
  1,
\end{equation}
so the horizon exit angular fluctuation is suppressed relative to the
minimal result by $\fa/\fI$. After the radial field returns to $\rho_0$, the
late-time axion coupling is still determined by $\fa$. This is the
large radius mechanism originally emphasized by Ref.~\cite{Linde:1991km}, and subsequently developed and reassessed in Refs.~\cite{Kearney:2016vqw, Kobayashi:2016qld, Graham:2025iwx}. When several PQ-charged fields acquire inflationary
background values, $\fI$ is instead the field-space norm of the corresponding
Goldstone direction. A large Higgs background in DFSZ models provides one
such multifield realization~\cite{Nakayama:2015pba}.

The post-inflationary radial evolution is essential. Oscillations of a
displaced PQ field can parametrically amplify radial and angular
fluctuations. If the fluctuations explore the full compact field space, PQ
symmetry can be restored nonthermally, regenerating strings and, once the QCD
potential becomes important, domain walls. Significant axion fluctuations
can also be produced even without complete restoration~\cite{Kofman:1995fi,Harigaya:2015hha,Kobayashi:2016qld,
Graham:2025iwx}. The radial field must therefore approach its late-time
minimum sufficiently smoothly, without excessive resonance, backreaction on
the inflationary or reheating background, or thermal PQ restoration.

In the very small self-coupling regime, radiative, thermal, and
curvature-induced corrections can substantially alter the PQ-field evolution
and its axion production~\cite{Kozow:2022whq}. When these effects remain
controlled, a sufficiently small PQ quartic can make the relaxation more
adiabatic and substantially enlarge the viable isocurvature parameter
space~\cite{Graham:2025iwx}. Finally, at fixed $\rho_I$ and $\fa$, increasing
$N_{\rm DW}$ reduces $\fI/\fa$. Achieving the same suppression then requires
a proportionally larger inflationary radial value. Nonthermal restoration is
especially problematic for $N_{\rm DW}>1$, because the resulting QCD domain
walls are stable in the absence of an additional bias.

\paragraph{Nonminimal couplings and curved field space.}
\label{subsec:nonminimal}
\label{subsec:curved}
More generally, $\fI$ is the field-space norm of the PQ angular direction.
When this direction is orthogonal to the remaining background fields
$\varphi^A$, the Einstein-frame metric may be written as
\begin{equation}
  \dd s_{\rm fs}^2
  =
  G_{AB}(\varphi)\,
  \dd\varphi^A\dd\varphi^B
  +
  F^2(\varphi)\,\dd\theta^2,
  \qquad
  \fI=F(\varphi_I).
  \label{eq:curvedmetric}
\end{equation}
The canonical evolving radius mechanism corresponds to the special case
$F=\rho/N_{\rm DW}$. A sufficiently small PQ quartic can then yield
$\fI\gg\fa$ during inflation while avoiding excessive axion production from
the subsequent radial evolution~\cite{Graham:2025iwx}.

Nonminimal gravitational couplings modify $F$ after the Weyl transformation
and can either weaken or strengthen the isocurvature bound, depending on the
couplings and inflationary trajectory~\cite{Rigouzzo:2025ycb}. The result also differs between metric and Palatini gravity. In minimal Higgs inflation, for example, the energy-dependent normalization strengthens the bound and restricts the
axion dark matter fraction to approximately~$\gamma_a\lesssim10^{-5}$~\cite{Rigouzzo:2025hza}.

Intrinsically curved PQ geometry can instead generate a time-dependent
effective mass for the angular mode, suppressing CMB-scale isocurvature while
producing a blue spectrum on smaller scales~\cite{Tadepalli:2026mdc}. If the
background evolves or the trajectory turns, however, the fluctuation and its
superhorizon transfer must be computed from the coupled multifield dynamics.
Substituting an instantaneous value of $F$ into the light-field formula is
not sufficient.

\subsection{Making the axion heavy during inflation}
\label{subsec:heavyaxion}

A temporary axion mass can evade the minimal light-field bound in two
physically distinct ways. If the mass is present while CMB modes leave the
horizon, it suppresses their production and subsequent superhorizon
amplitude. If it arises only after inflation, it instead damps an already
existing perturbation and modifies its transfer into late-time
isocurvature.

For approximately constant $\HI$, $\fI$, and $m_{a,I}$ in quasi-de Sitter
space, a canonically normalized axion fluctuation has
\begin{equation}
  \nu
  =
  \sqrt{
    \frac{9}{4}
    -
    \frac{m_{a,I}^2}{\HI^2}
  },
  \qquad
  |\delta a_k|
  \propto
  a^{-3/2+\operatorname{Re}\nu}
  \qquad
  (k\ll a\HI).
  \label{eq:heavy-superhorizon}
\end{equation}
Thus, for $0<m_{a,I}<3\HI/2$, a CMB mode is damped outside the horizon by
$\exp[-(3/2-\nu)\Delta N]$. For $m_{a,I}>3\HI/2$, $\nu$ is imaginary and
the superhorizon envelope decays as $a^{-3/2}$. Particle production is
additionally Boltzmann suppressed when $m_{a,I}\gg\HI$. The magnitude of the
suppression therefore depends on both the mass and the duration for which it
is active.

\paragraph{Early confinement and small instantons.}
\label{subsec:strong}
If the gauge coupling, colored thresholds, or Yukawa couplings depend on the
inflationary background, the effective topological susceptibility can greatly
exceed its standard-QCD value. Define its inflationary value by
\begin{equation}
  \chi_I
  \equiv
  \left.
  \frac{\partial^2 V_{\rm strong}(\theta)}
       {\partial\theta^2}
  \right|_{\theta=\theta_{\rm min},\,I},
  \qquad
  m_{a,I}^2
  =
  \frac{\chi_I}{\fI^2},
  \label{eq:chiI}
\end{equation}
where $V_{\rm strong}(\theta)$ is the $\theta$-dependent vacuum energy of the
relevant confining sector. If $m_{a,I}$ is sufficiently large relative to
$\HI$ for a sufficiently long interval, the CMB-scale axion fluctuation is
strongly suppressed.

Realizations include modulus-dependent gauge couplings, large
inflationary Higgs expectation values, ultraviolet small instantons, and a
temporarily confining unified gauge theory~\cite{Dvali:1995ce, Jeong:2013xta, Choi:2015zra, Buen-Abad:2019uoc, Dvali:2026ceb}. Direct inflaton--gluon couplings can likewise drive early QCD confinement. Recent analyses show that maintaining controlled confinement evolution favors plateau-like inflationary
backgrounds~\cite{Freese:2026xax, Sfakianakis:2026rge}. Because the associated strong-sector vacuum energy depends on the inflationary background, its effects on the inflaton slope, scalar tilt, and reheating dynamics must be
included. If the enhanced susceptibility becomes important only after CMB
horizon exit, it modifies the abundance and perturbation transfer rather
than the primordial fluctuation itself.

\paragraph{Explicit breaking and topological masses.}
\label{subsec:explicit}
Inflation-dependent PQ breaking can generate a temporary potential for the
dimensionless axion angle,
\begin{equation}
  V_I(\theta)
  =
  \Lambda_I^4
  \left[
    1-
    \cos\!\left(
      N_I\theta+\delta_I
    \right)
  \right],
  \qquad
  m_{a,I}^2
  =
  \left.
  \frac{1}{\fI^2}
  \frac{\partial^2V_I}{\partial\theta^2}
  \right|_{\rm min}
  \simeq
  \frac{N_I^2\Lambda_I^4}{\fI^2}.
  \label{eq:explicit-mass}
\end{equation}
Such a potential can arise from inflation-dependent operators, discrete gauge
symmetries, or PQ-breaking nonminimal couplings~\cite{Higaki:2014ooa, Kearney:2016vqw, Berbig:2024ufe, Kawasaki:2026jen}. BF-type or three-form couplings provide a topological realization in which a quantized mass parameter is removed through flux discharge~\cite{Chakraborty:2025lyp}.

\paragraph{The Witten effect and post-inflationary damping.}
\label{subsec:witten}
An axion coupled to a hidden Abelian sector acquires a monopole-induced
potential because magnetic monopoles carry axion-dependent electric charge~\cite{Witten:1979ey}. For a coupling normalized by an
effective scale $f_H$, the induced mass scales as
\begin{equation}
  m_{a,M}^2
  \sim
  \frac{\alpha_H}{16\pi^2}
  \frac{n_M}{r_c f_H^2},
  \label{eq:wittenmass}
\end{equation}
where $n_M$ and $r_c$ are the monopole density and core radius. The numerical
coefficient and screening corrections are model dependent. In the standard
cosmological implementations, monopoles are produced after inflation and the
induced mass triggers early axion relaxation or oscillations, damping the
pre-existing isocurvature perturbation and modifying the relic abundance~\cite{Kawasaki:2015lpf, Nomura:2015xil}. In two-stage hidden sector breaking, monopoles can subsequently become confined by strings and annihilate, removing the temporary potential~\cite{Banerjee:2024ykz}.

All temporary-mass mechanisms must control the transition to ordinary QCD.
The early minimum must be aligned with the QCD minimum or evolve into it in a
controlled manner. Otherwise the turn-off can regenerate misalignment, axion
quanta, spatial fluctuations, or defects. Controlled early relaxation and
moving minima can instead be used to select the late-time misalignment angle~\cite{Co:2018mho, Buen-Abad:2019uoc}. Any explicit breaking that
survives at late times must also satisfy the axion-quality and strong-CP
constraints.

\subsection{Changing the abundance and perturbation transfer}
\label{subsec:transfer}
Standard misalignment reduces the late axion-abundance perturbation to the
single-variable response $\Gresp(\thetai)=\partial\ln\Omegaa/\partial\thetai$. More general production histories depend on several fluctuating initial data $X_A$. At linear order,
\begin{equation}
  \delta\ln\Omegaa
  =
  \sum_A
  \mathcal G_A\,\delta X_A,
  \qquad
  \mathcal G_A
  \equiv
  \frac{\partial\ln\Omegaa}{\partial X_A},
  \label{eq:multivariable-response}
\end{equation}
where the derivatives are evaluated on the homogeneous background. The
primordial covariance matrix of the $X_A$ and their subsequent transfer must
therefore be evolved together. A single response function of $\thetai$ no
longer suffices.

\paragraph{Kinetic misalignment.}
A rotating PQ condensate carries an approximately conserved comoving PQ
charge, so the abundance depends on its initial charge, or angular velocity,
as well as on the initial angle~\cite{Co:2019jts}. Inflationary fluctuations of the angular and radial fields, together with post-inflationary parametric resonance or fragmentation, must then be propagated into the final abundance~\cite{Co:2020dya}. Kinetic misalignment therefore changes
the abundance response but does not generically eliminate isocurvature.

\paragraph{Axion mixing.}
A temperature-dependent level crossing can resonantly convert QCD axions into
a hidden axion. Efficient conversion suppresses the QCD axion abundance,
while incomplete conversion caused by a mild violation of adiabaticity can
suppress the CDM isocurvature power without similarly suppressing its
non-Gaussianity~\cite{Kitajima:2014xla}. The observable constraint must
be applied to the total CDM entropy perturbation, including any
cosmologically significant hidden-axion component.

\paragraph{Entropy release and dissipative evolution.}
A late-decaying saxion, modulus, or other field can inject entropy and dilute
the axion abundance~\cite{Steinhardt:1983ia,Kawasaki:1995vt,Kawasaki:2013ae}.
If the entropy injection is spatially homogeneous, it changes the mean
$n_a/s$ but leaves the intrinsic perturbation $\delta\ln(n_a/s)$ unchanged. The observable CDM isocurvature is reduced only
insofar as the axion dark matter fraction decreases. If the axion abundance
is subsequently retuned, or if the decaying sector is itself perturbed, the
full abundance response and transfer must be recomputed.

Tachyonic production of dark photons can dissipate the axion condensate and
modify both its relic abundance and perturbation spectrum~\cite{Agrawal:2017eqm}. For generic scalar dark matter, thermal misalignment can instead suppress isocurvature through a late-time phase offset between the homogeneous condensate and its superhorizon perturbation~\cite{Batell:2026sml}. Applying the latter mechanism to the QCD axion would require an explicit finite-temperature realization consistent with axion quality, the shift symmetry, and the assumed PQ history.

\subsection{Changing the PQ history or the primordial initial condition}
\label{subsec:PQhistory}
If PQ symmetry is restored after inflation and breaks again later, the
inflationary angular mode is erased. The cosmology is then post-inflationary:
the angle varies among Hubble patches, strings form at the PQ transition, and
domain walls appear when the axion potential becomes important. The abundance
and small-scale density field must be computed including defects~\cite{Gorghetto:2018myk, Buschmann:2021sdq, Kawasaki:2014sqa}. Restoration can also be nonthermal. For example, an inflaton--PQ interaction can keep the symmetry unbroken during inflation even when $\fa>\HI$, again replacing the
pre-inflationary initial condition with a post-inflationary one~\cite{Bao:2022hsg}.

At sufficiently low $\HI$, another regime becomes possible. If inflation
lasts long enough for the axion to approach its de Sitter equilibrium
distribution, the local angle is not an arbitrary fixed parameter. Its
abundance and isocurvature must instead be averaged over the stochastic
distribution~\cite{Graham:2018jyp,Takahashi:2018tdu}. Additional light sector
isocurvature sources can also become relevant in low-scale inflation~\cite{Caputo:2023ikd}.

Finally, the mapping from primordial axion fluctuations to the CMB can differ
from the scale-invariant, uncorrelated, all dark matter benchmark. A
subdominant axion fraction suppresses the total matter-isocurvature amplitude;
correlated or compensated modes change the CMB transfer functions; and a blue
spectrum weakens the connection to the pivot constraint while introducing
small-scale limits, including spectral-distortion constraints~\cite{Kasuya:2009up, Chluba:2013dna}. These cases require the appropriate
primordial spectra and correlations to be propagated through a dedicated CMB
likelihood rather than represented by a rescaled flat-CDI limit~\cite{Petretti:2026ayw}.

The minimal bound is therefore not a no-go theorem, but each genuine evasion
replaces one of its assumptions with additional dynamics. A complete model
must identify which ingredient of Eq.~\eqref{eq:masterboundPII} is changed,
derive the corresponding fluctuation and transfer functions, and verify that
the subsequent evolution does not regenerate isocurvature, defects,
misalignment, or a strong-CP phase.

\section{Discussion and conclusions}
\label{sec:conclusions}
For a light axion producing a linear, uncorrelated CDI mode, the general
pivot-scale ceiling is given by Eq.~\eqref{eq:masterboundPII}. Its strength
depends on the inflationary canonical normalization $\fI$, the axion
dark-matter fraction $\gamma_a$, the perturbation transfer $\Ttheta$, and the
abundance response $\partial_{\thetai}\ln\Omegaa$. In the minimal
pre-inflationary benchmark, where $\fI=\fa\leq\Mpl$, $\Ttheta=1$, and axions
constitute all of the dark matter, the dedicated likelihood analysis of
Ref.~\cite{Petretti:2026ayw} gives $\HI<1.25\times10^{10}\GeV$ and $\HI<2.2\times10^7\GeV~(\thetai=1)$, at 95\% confidence. The first and least restrictive ceiling is attained only
near $\fa\simeq\Mpl$, where the all dark matter abundance condition requires
a very small initial angle. Near the hilltop, the anharmonic abundance
response grows rapidly and can lower the allowed scale by many further orders
of magnitude.

The hilltop result cannot be extrapolated arbitrarily close to
$\thetai=\pi$. Once
$|\Gresp(\thetai)|\,\delta\thetai\not\ll1$ or
$\delta\thetai\not\ll\pi-\thetai$, the linear Gaussian treatment must be
replaced by a nonlinear stochastic calculation. More generally, there is no
model-independent numerical ceiling without specifying $\fI/\fa$, the axion
fraction, the relic and transfer histories, and the allowed ultraviolet range
of $\fa$. For the scale-invariant CDI spectrum considered here, adding ACT or
SPT data does not strengthen the Planck amplitude limit. P-ACT gives the
weakest of the three constraints considered.

The main implication for inflation is not merely that $\HI$ must be small.
In canonical cold single-field slow roll, the scalar amplitude and red tilt
require $\epsilon_V \simeq 10^{-16}\text{--}10^{-10}~(\HI=10^7\text{--}10^{10}\GeV)$. Thus $\epsilon_V/|\eta_V|$ can be as small as $10^{-14}$: the potential must have an exceptionally small local slope while retaining percent-level negative curvature. This axion-conditioned slope--curvature hierarchy is the central model-building consequence of the minimal bound.

Assuming Einstein gravity and the standard vacuum tensor spectrum,
Eq.~\eqref{eq:axiontensorPII} implies an unobservably small tensor amplitude
throughout the minimal parameter space. The corresponding axion-conditioned
Lyth estimate, Eq.~\eqref{eq:axionlythPII}, restricts the inflaton displacement
across the observable CMB window to
$|\Delta\phi_{\rm CMB}|/\Mpl\lesssim10^{-7}\text{--}10^{-4}$, provided the
tensor amplitude does not vary sharply over this interval. Whereas the
ordinary Lyth argument~\cite{Lyth:1996im,Efstathiou:2005tq} associates an
observable tensor signal with substantial field motion, the minimal
pre-inflationary axion selects the opposite regime: the inflaton is nearly
frozen while observable modes leave the horizon. This remains a local
statement and does not bound the total field excursion outside the CMB
window.

Reheating closes the logical loop. For the post-inflationary histories
considered here, lowering the inflationary scale reduces the preferred value
of $\Ns$. Thermal PQ non-restoration may additionally require a lower
reheating or maximum temperature, or sufficiently weak couplings between the
PQ sector and the plasma. When it enforces delayed matter-like reheating,
$\Ns$ decreases further. Nonthermal restoration instead requires control of
the radial PQ dynamics and does not follow from the reheating temperature
alone. Universal plateau and pure $p>2$ hilltop models therefore become
redder precisely where the low inflationary scale and delayed reheating push
the theory. The larger scalar tilt preferred by ACT strengthens this
model-building filter even though ACT does not tighten the direct
scale-invariant isocurvature-amplitude constraint.

The inflationary survey points to a simple structural criterion: a robust
canonical low-scale construction should permit largely independent control
of the vacuum energy $V_0$, the pivot-scale slope $V_*'$, the curvature
$V_*''$, the end of inflation, and the reheating and PQ histories.
Vacuum-dominated hybrid inflation realizes this separation most directly:
the vacuum energy fixes the scale, the local derivatives determine the scalar
spectrum, and a waterfall field ends inflation. Running-mass,
quadratic hilltop, and inflection-point models can realize the same
phenomenology, but typically with greater sensitivity to ultraviolet
thresholds, radiative corrections, relations among local coefficients, or
initial conditions. Spectator, noncanonical, and warm inflation mechanisms can modify the usual relation between the scalar amplitude and the inflaton potential, but they do not remove the direct axion bound on $\HI$ unless the axion sector or its subsequent transfer is also changed.

The constraint can therefore be accommodated in three broad ways: lower the
inflationary scale, suppress the primordial axion fluctuation, or alter its
relic abundance and conversion into late-time matter isocurvature. Examples
of the latter two possibilities include early confinement, controlled
explicit PQ breaking, time-dependent canonical normalization, nonminimal
gravitational couplings, curved PQ field space, post-inflationary PQ breaking,
axion mixing, monopole-induced damping, entropy release, and other
nonstandard relic histories. These mechanisms are most clearly classified
by the assumption entering Eq.~\eqref{eq:masterboundPII} that they principally
relax. Several can reopen high-scale inflationary regimes excluded by the
minimal light axion scenario, but each introduces additional requirements
involving alignment, backreaction, defect production, field-space stability,
reheating, or perturbation transfer.

The observational consequences are correspondingly sharp. A future
primordial vacuum-tensor detection corresponding, under the standard tensor
relation, to $\HI>1.25\times10^{10}\GeV$ would exclude the minimal scenario
in which a light QCD axion with $\fI=\fa\leq\Mpl$ constitutes all of the dark
matter. It would instead require a modified inflationary axion normalization
or mass, a subdominant axion abundance, a post-inflationary PQ transition, or
another nonminimal history. Conversely, if the scalar tilt settles near the
representative ACT DR6 value $n_s\simeq0.974$, with no detectable tensors or
large running, the favored interpretations would include low-scale models
with independently adjustable curvature, such as hybrid, running-mass,
quadratic hilltop, or inflection-point inflation, or an axion mechanism that
suppresses the primordial entropy mode and permits a higher inflationary
scale.

Taken together, the direct isocurvature limit, scalar tilt, reheating,
tensor ceiling, and restricted CMB-scale field motion turn the
pre-inflationary QCD axion from a one-line bound on $\HI$ into a falsifiable
statement about the architecture of inflation. In the minimal scenario,
inflation must be exceptionally low scale, locally flat in slope but not in
curvature, and compatible with a PQ-safe post-inflationary history. If future
data contradict any of these conclusions, the nature of the discrepancy will
identify which assumption about the axion or inflationary dynamics must be
relaxed.

\section*{Data availability statement}
No new data were generated or analyzed in this study.

\section*{Acknowledgments}
I thank Marcos A.~G.~Garc\'\i a and Santiago Ag\"u\'\i{} Salcedo for useful
discussions. The work of S.V. was supported by the Kavli Institute for
Cosmological Physics at the University of Chicago.

\addcontentsline{toc}{section}{References}
\bibliographystyle{JHEP}
\bibliography{ref} 

\end{document}